\documentclass[prd,twocolumn,showpacs,preprintnumbers,amsmath,amssymb,nofootinbib]{revtex4-1}
\pdfoutput=1
\usepackage{graphicx}
\usepackage{hyperref}

\def\be{\begin{equation}}
\def\ee{\end{equation}}
\def\bea{\begin{eqnarray}}
\def\eea{\end{eqnarray}}
\def\ben{\begin{equation}}
\def\een{\end{equation}}
\def\bean{\begin{eqnarray}}
\def\eean{\end{eqnarray}}
\def\mbf{\mathbf}

\numberwithin{equation}{section}
\def\line{\vspace{3mm}}

\newcommand{\one}{\leavevmode\hbox{\small1\kern-3.8pt\normalsize1}}
\newcommand{\im}{\mathrm{i}}
\newcommand{\Vext}{V_\mathrm{ext}}
\newcommand{\x}{\mathbf{x}}
\renewcommand{\v}{\mathbf{v}}
\newcommand{\bnabla}{\mathbf{\nabla}}

\begin{document}

\title{Analogue Cosmological Particle Creation:\\ Quantum Correlations in Expanding Bose Einstein Condensates}

 \author{Angus Prain$^{1}$}
 \email{aprain@sissa.it}
 \author{Serena Fagnocchi$^{2}$}
 \email{fagnocchi@nottingham.ac.uk}
 \author{Stefano Liberati$^{1}$}
 \email{liberati@sissa.it}
\affiliation{${}^1$SISSA,  Via Bonomea 265, 34136 Trieste, Italy and \\ INFN sezione di Trieste, Via Valerio 2, 34127 Trieste, Italy}
\affiliation{${}^2$School of Physics \& Astronomy,  University of Nottingham, University Park, Nottingham NG7 2RD, UK}
%
%
 \date{\today}
%
%
%
\begin{abstract}
{We investigate the structure of quantum correlations in an expanding Bose Einstein Condensate (BEC) through the analogue gravity
framework. We consider both a 3+1 isotropically expanding BEC as well as the experimentally relevant case of an elongated, effectively 1+1 dimensional, expanding condensate. In this case we include the effects of inhomogeneities in the condensate, a feature rarely included in the analogue gravity literature.
In both cases we link the BEC expansion to a simple model for an expanding spacetime and then study the correlation structure numerically and analytically (in suitable approximations).  We also discuss the expected strength of such correlation patterns and experimentally feasible BEC systems in which these effects might be detected in the near future.}
\end{abstract}
\pacs{03.75.Kk, 05.30.Jp, 04.62.+v, 04.70.Dy}

\maketitle

\section{Introduction}

Within the framework of quantum field theory in curved spacetime where fields
 propagate on a fixed  background, some interesting -- and sometimes surprising -- quantum instabilities are predicted.  Among them a crucial role has been played by the so called  \emph{cosmological particle creation}: a quantum emission of particles as a consequence of an
 expansion of spacetime \cite{Parker:1969au, Birrell:1982ix}. Being associated with wave propagation over a dynamical space and not (when backreaction is neglected) to a property of the Einstein equations, it is not peculiar to the gravitational framework but appears as a quantum field non-equilibrium process in a diverse range of physical systems. This is a feature shared with other semiclassical emissions,  for example  Hawking radiation \cite{Hawking:1974rv}. For this reason, recently there has been a growing interest in studying such non-trivial quantum effects in a laboratory setting with condensed matter systems, a field of research that goes under the general name of ``analogue models of gravity" (see e.g.~\cite{Barcelo:2005fc} for an extensive review).

In particular, the propagation of linearized acoustic perturbations
on an inviscid and irrotational fluid can be shown to be described
by the same equation of motion as that which describes the propagation of a
scalar field on a curved spacetime. Of course such ideal systems are
not readily available in nature but it is possible to produce
systems which, to a very good degree of approximation, do actually
behave as perfect fluids. In recent years a special role in this
sense has been played by Bose-Einstein condensates (BEC), the basic
observation being in this case that phase perturbations of the wave
function describing a weakly interacting BEC are, in a certain
limit, described by a massless quantum scalar field propagating on a
non-trivially curved background Lorenzian space-time
\cite{garay-2000-85, Barcelo:2000tg, Barcelo:2005fc}.

A BEC can therefore be used to simulate special gravitational
backgrounds and the associated particle production such as Hawking
radiation of black holes \cite{Parentani:2010bn,
Balbinot:2007de,Carusotto:2008ep, Schutzhold:2010ig}, or the
cosmological emission mentioned above that we are concerned with in
this paper. For example, releasing the trapping potential of a BEC
appropriately \cite{Gibbons,Uhlmann:2005hf, Fischer:2004hr,
fedichev-2004-69, PhysRevD.69.064021, PhysRevLett.99.201301,
barcelo-2003-12} or modulating external parameters of the
experimental setting \cite{Barcelo:2003wu,Jain:2007gg,Visser:2007du,Weinfurtner:2008if,Carusotto:2009re}, it is
possible to induce the necessary dynamics which give rise to
interesting background geometries.  Due to the non-trivial
time evolution, the system undergoes a non-equilibrium phase to
which are associated quantum emissions whose spectrum strictly
depends on the parameters modulating the evolution. This opens up
the possibility of experimentally accessing new kinds of phenomena
directly involving the quantum nature of cold atom systems as well
as providing additional insights on the general mechanisms of such
emissions in the gravitational case.


Cast in a different light, this possibility of describing terrestrial experiments using a curved spacetime analogue provides one with a new set of tools to tackle difficult problems in condensed matter physics and is able to suggest new and interesting non-equilibrium quantum effects.  Moreover, a successful application of these alternative techniques and consequent observation of the effects predicted would serve as further theoretical support for the validity of the semiclassical theory.
\line \\
The issue of how to practically observe these quantum emissions and verify their origin is a crucial point in the condensed matter literature. In particular there has been recently a large interest in the calculation and  measurement of correlation patterns in expanding condensates to access these quantum effects (for example see \cite{PhysRevA.81.031610, PhysRevA.80.033604, PhysRevA.67.043603}).  In \cite{Balbinot:2007de} it has been proposed that evidence for analogue Hawking emission in a sonic black hole manifested in a BEC might be solicited by measuring the two point correlation structure associated with the quantized density or phase perturbation fields. The idea is that the correlation pattern should display a peculiar signature of the type of emission expected from black holes. Following the same line of reasoning, we apply this idea to analogue cosmological emission and attempt to isolate a cosmological signal in the correlation pattern in an expanding BEC.

In particular this article is concerned with the experimentally accessible correlations between the phase and density perturbations in a BEC which are born during the expansion of the underlying gas both in the widely studied effectively 1+1 dimensional ``cigar" shaped geometry and in a homogeneous expanding BEC.  Such BEC systems are of theoretical interest in their own right and correlation measurements are experimentally accessible with the current technology \cite{PhysRevA.81.031610, PhysRevA.80.033604, PhysRevA.67.043603}. Therefore it is possible that these systems will provide in the near future an arena within which to experimentally address interesting and new non-equilibrium quantum effects while at the same time shed light on the physics of quantum fields in expanding spacetime without using telescopes.

In the first section we collect and contextualize some results relevant for the study of correlations in an expanding BEC, reviewing how the analogy is applied to BEC and the dynamics of an expanding BEC. In the second section we compute the correlation structure for a massless scalar field in a particular expanding background spacetime in 3+1D and show how to interpret such a result in terms of both an isotropically expanding 3+1D condensate and a BEC with a time dependent scattering length.  We shall develop a general formalism for probing the effect of particle production (Bogoliubov coefficients) on correlations as embodied in the equal-time Wightman functions.
In this context we shall derive results which are completely generic to particle production in any spatially translation invariant system, which can be applied to standard cosmologies as much as to ``analogue spacetimes".
In the third and final section which represents the main result of this article using both analytic and numerical methods we discuss the structure of correlations in an anisotropically expanding, effectively $1+1$-dimensional, BEC.  In particular we shall derive the general, finite size, $1+1$-dimensional, equal time Wightman function relevant for the ``cigar" shaped expanding BEC presently realized in experiments.  We conclude with some proposals for improving the observability in practice of the correlation signal.

\section{Background}

\subsection{A BEC as an analogue spacetime}

In this article we consider a weakly interacting BEC trapped in an attractive harmonic potential.  Given the by now standard usage of this system as an analogue model of gravity we limit ourselves to a summarry of the salient equations which we shall need in what follows, directing the reader to the extant literature for further reading \cite{garay-2000-85, Barcelo:2000tg, Barcelo:2005fc}.

In the dilute gas approximation, one can describe a Bose gas through a quantum field ${\widehat \Psi}$ satisfying
\be
\im\hbar \; \frac{\partial }{\partial t} {\widehat \Psi} =
\left(
  - \frac{\hbar^2}{2m} \nabla^2 + \Vext(\x)
  +g(\bar{a})\;{\widehat \Psi}^{\dagger}{\widehat \Psi}
\right){\widehat \Psi}.
\label{LG-EOM}
\ee
Here $m$ is the mass of the constituents,  $g$ parametrizes the strength of the interactions between pairs of bosons in the gas and $\bar{a}$ is the scattering length of the atoms in an s-wave approximation.  The former can be re-expressed in terms of the scattering length as
\begin{equation}
g(\bar{a}) = \frac{4\pi \bar{a} \hbar^2}{m}.
\end{equation}

Adopting a mean field approximation, the quantum field can be
separated into a macroscopic (classical) condensate (called the wave
function of the condensate) and a fluctuation:
${\widehat\Psi}=\psi+{\widehat \varphi}$, with $\langle {\widehat
\Psi} \rangle=\psi $. Then, by adopting the self-consistent mean
field approximation and neglecting backreaction effects of the
quantum fluctuations on the condensate, one obtains the
Gross--Pitaevskii (GP) equation of motion for the wave function of
the condensate
\be
i\hbar\frac{\partial\psi}{\partial t}=-\frac{\hbar^2}{2m}\nabla^2\psi+V_{\text{ext}}\psi+g(\bar{a})|\psi|^2\psi.  \label{GP}
\ee
Now, adopting the Madelung representation for the wave function of the condensate \cite{Dalfovo:1999zz}
\begin{equation}
\psi(t,\x)=\sqrt{\rho(t,\x)} \; \exp[-\im\theta(t,\x)/\hbar],
\end{equation}
where $\rho=\psi^*\psi$ is the condensate density,  and defining an
irrotational ``velocity field'' by $\v:={\bnabla\theta}/{m}$, the
Gross--Pitaevskii equation can be rewritten as a continuity equation
plus an Euler equation:
\begin{align}
& \quad\quad\frac{\partial}{\partial t}\rho+\bnabla\cdot({\rho \v})=0,
\label{E:continuity}\\
m\frac{\partial}{\partial t}\v+\bnabla&\left(\frac{mv^2}{2}+
V_\mathrm{ext}(t,\x)+g \rho- \frac{\hbar^2}{2m}
\frac{\nabla^{2}\sqrt{\rho}}{\sqrt{\rho}} \right)=0.
\label{E:Euler1}
\end{align}
These equations are completely equivalent to those of an irrotational and inviscid fluid apart from the presence of the so-called quantum
potential
\begin{equation}
V_\mathrm{quantum}=-\hbar^2\nabla^{2}\sqrt{\rho}/(2m\sqrt{\rho}),
\end{equation}
which has the dimensions of an energy. It is then clear that when small gradients of the background density are involved, this term can be safely neglected and a hydrodynamical approximation (also known  as the Thomas Fermi (TF) approximation) holds.

We are concerned with the quantum excitations on top of the BEC
background solution of the GP equation (or hydrodynamic
equivalents).  Inserting the mean field ansatz in Eq.~\eqref{LG-EOM}
and again neglecting backreaction, one obtains the equation for the
operator valued perturbation
\begin{align}
i \hbar \; \frac{\partial }{\partial t} {\widehat \varphi}(t,\x)
&= \left( - \frac{\hbar^2}{2m} \nabla^2  +  V_\mathrm{ext}(\x)   +2g \rho\right) {\widehat \varphi} (t,\x) \nonumber\\
&\hspace{25mm}+  g\, \rho\, {\widehat \varphi}^{\dagger}(t,\x).
\label{quantum-field}
\end{align}
%
The elementary excitations of this dynamical system are described by variables obtained by  a Bogoliubov transformation from the fundamental field.  This can be also shortcut by adopting the so called quantum acoustic representation \cite{Barcelo:2005fc}
\begin{eqnarray}
\widehat
\varphi(t,\x)=&& e^{-\im \theta/\hbar} \left({1 \over 2
\sqrt{n_c}} \; \widehat \rho_1 - \im \; {\sqrt{\rho} \over \hbar} \;\widehat
\theta_1\right),
\label{representation-change}
\end{eqnarray}
where
$\widehat \rho_1,\widehat\theta_1$ are real quantum fields which describe the excitations in density and phase respectively.
Inserting the above ansatz in Eq.~\eqref{quantum-field} one again obtains a pair of hydrodynamic-like equations for the quantum excitations
\begin{align}
&\partial_t \widehat \rho_1 + {1\over m}
\bnabla\cdot\left( \rho_1 \; \bnabla \theta + \rho \; \bnabla \widehat
\theta_1 \right) = 0, \label{pt1}
\\
 &\partial_t \widehat \theta_1   +
\frac{\bnabla \theta \cdot \bnabla \widehat \theta_1}{m}
+ g(\bar{a}) \; \rho_1 - {\hbar^2\over2 m}\; D_2 \widehat \rho_1 = 0,
\label{pt2}
\end{align}
where $D_2$ represents a second-order differential operator derived
from the linearization of the quantum potential
\cite{Barcelo:2005fc}. In a long wavelength approximation this can
be safely neglected and one can then show (in complete analogy with
perfect fluid systems) that the perturbations $\theta_1$ do satisfy
a second order equation of a Klein Gordon type for a massless scalar field propagating on a curved
background geometry described by the metric
\be
ds^2=\sqrt{\frac{\rho}{gm}}\left[-(c_s^2-v^2)dt^2-2v_idx^i\,dt+d\mbf{x}^2\right]. \label{back}
\ee
Here $c_s^2(t,\mbf{x})=g\rho(t,\mbf{x})/m$ is the squared local
sound speed of the perturbations and $\mbf{v}(t, \mbf{x})$ is the
background flow velocity of the BEC introduced above
\cite{Visser:1993ub,garay-2000-85}.  It is then clear that in this
long wavelength regime the quasi-particles satisfy a relativistic,
phononic dispersion relation $\omega^2=c^2_s k^2$ (with $\omega$
and $k$ equal respectively to the energy and wavenumber of the
quasi-particle).  Alternatively one can adopt an eikonal (short
wavelength) approximation and show that the complete dispersion
relation is indeed the well known Bogoliubov dispersion given by
$\omega_{Bog}^2=c_s^2k^2+\hbar k^4/(2m)$.  It is evident at this
point that in this context the gravitational analogy is valid only
in the long wavelength approximation. {More specifically,  for
wavelengths of the order of the healing length
$\xi=\hbar/\sqrt{mg\rho}$, when the quartic term
becomes comparable to the quadratic
one in $\omega_{Bog}^2(k)$, a geometrical description is no longer available.}

\subsection{Scaling solution}\label{scaling}

Condensates undergoing expansion or contraction as a result of a
time varying confining harmonic potential admit a dynamic
generalization to the static TF approximation for the GP equation
known as the scaling solution \cite{Castin:1996zz,
2009arXiv0912.2744G}.  In this case the external potential is of
harmonic type with time varying trapping frequencies
\be
 V_{\text{ext}}=\frac{1}{2}\sum_im\omega_i^2(t)x_i^2.
 \ee
 Here the density undergoes a simple scaling transformation in each spatial dimension $x_i$ parametrized by the so-called scaling functions $\lambda_i(t)$ as
\be
\rho(t,\mbf{x})=\frac{\rho_s(x_i/\lambda_i(t))}{\prod_i\lambda_i(t)}\label{eq:scale},
\ee
where $\rho_s$ is the static TF  ground state solution\footnote{Note that the TF approximation is not a good approximation near the boundary of the condensate where the density drops to zero.  The condensate boundary, here given by the parabola sharply dropping to zero, in the full solution will be smooth.}
\begin{align}
\rho_s&=\rho_0 \left(1-\sum_i\tilde{\omega}_ix_i^2\right)\quad \text{if} \quad \sum_i\tilde{\omega}_ix_i^2<1 \, ,\\
\rho_s &=0  \quad \text{otherwise},
\label{sol}
\end{align}
where $\tilde{\omega}_i={m\omega_i^2}/{2\mu}$ ($\mu$ being the chemical potential) and  $\rho_0$ is the maximum density at the centre  of the BEC.

In the static TF approximation the density and chemical potential can be expressed algebraically in terms of the physical parameters  as
\be
\rho_0=\frac{\mu}{g(\bar{a})}=\frac{1}{8\pi}\left(\frac{15 m^3 N \prod_i\omega_i}{\bar{a}^{3/2}\hbar^3}\right)^{2/5}
\ee
and
\be
\mu=\frac{1}{2}\hbar{\varpi}\left(15 N\bar{a}\sqrt{\frac{m\varpi}{\hbar}}\right)^{2/5}.
\ee
respectively, where $N$ is the number of condensed bosons.  Here we have introduced the notation $\varpi=\left(\prod_i \omega_i\right)^{1/d}$ in $d$ spatial dimensions, the geometric mean of the trapping frequencies. For the phase, the scaling transformation takes the form
\be
\theta(t,\mbf{x})=\frac{1}{2}m\sum_ix_i^2\dot{\lambda}_i\lambda_i,
\ee
where $\dot{}$ represents the lab time derivative $\partial_t$.

This scaling induces a background flow velocity profile in the condensate given by
\be
\mbf{v}(t,\mbf{x})=\frac{\dot{\lambda}(t)}{\lambda(t)}\,\mbf{x}.
\ee
With this velocity and the evolution of the density given in \eqref{eq:scale} one can start studying the analogue geometry associated with such a scaling background solution using the metric \eqref{back}.

Of particular experimental interest is the ``cigar'' shaped condensate possessing one elongated dimension (``$||$'') and two tightly trapped orthogonal dimensions (``$\perp$''). In what follows, unless otherwise stated, we work with the cigar geometry.  Since our main focus will be on such elongated condensates, we list here the following realistic values for the physical parameters describing the cigar BEC
\bea
&&\omega_{||}=10^2\text{Hz}, \quad \omega_\perp=10^3\text{Hz} \\
&&m= 1.57\times10^{-25} \,\text{kg    (Rubidium $87$) }  \\
&&\bar{a}=42\,\text{Bohr}=2.221\times10^{-9}\;\text{m}  \\
&& N=10^5.
\eea
These are respectively the trapping frequencies, atomic mass, scattering length and number of condensed atoms.  Assuming these numbers, the following derived quantities are obtained
\bea
&& L\approx5.45\times 10^{-5}m\\
&&\mu\approx 5.8\times10^{-31}\,\,\text{kgm$^2$s$^{-2}$}\approx 5500\hbar\, \text{s}^{-1}\\
&& \rho_0\approx 2.9\times 10^{20}\, \text{m}^{-3} \\
&& c_{s,0}\approx 2 \,\,\text{mms$^{-1}$}
\eea
corresponding to the transverse condensate size (where, in the TF approximation the density drops to zero), the (static) chemical potential, the maximum condensate density and the maximum initial sound speed respectively.

For time dependent harmonic traps the scaling functions $\lambda_i(t)$ are not free parameters but are constrained to be solutions to the auxiliary equations of motion
\be
\ddot{\lambda}_i(t)+\omega_i^2(t)\lambda_i(t)=\frac{\omega_i^2(0)}{\lambda_i(t)\prod_j\lambda_j(t)} \label{scalingf}
\ee
with boundary conditions $\lambda_i(0)=1$ and $\dot \lambda_i(0)=0$, determined by a self consistency condition on the dynamics of the expanding condensate \cite{Castin:1996zz}.

Starting from such a cigar shaped initial geometry, let us consider
the case when the ``$||$'' potential is suddenly switched off while
keeping the ``$\perp$'' potentials constant. A similar problem has
been studied in \cite{PhysRevA.80.033604} and observed in
\cite{PhysRevA.81.031610} where the authors instead  consider a BEC
freely expanding in all the three spatial dimensions after
completely releasing the trapping potential of an initially cigar
shaped condensate. In these analyses, due to the  expansion in all
three dimensions, the density drops very quickly and the gas becomes
collisionless signalling the failure of the hydrodynamical description shortly after
the release of the trap.  In the case we consider here instead,
the perpendicular trapping is kept tight and the density drops more
slowly and the condensate remains longer in a collisional regime
where the spacetime analogy pertains.

\begin{figure}
\begin{center}
\includegraphics[scale=0.4]{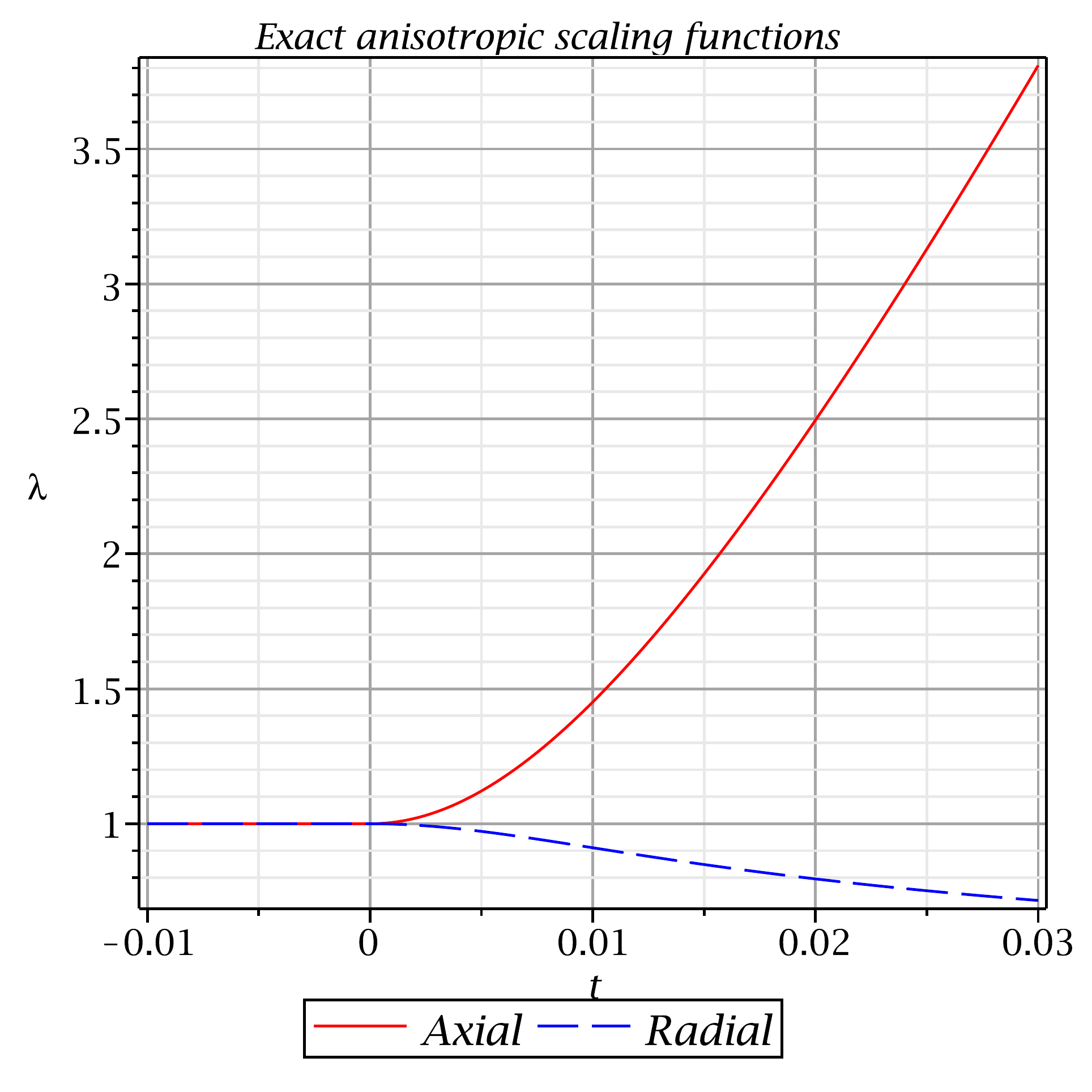}
\caption{The solution of the scaling problem for an anisotropic expansion of an initially cigar shaped configuration for both the radial/orthogonal (blue-dashed line) and axial/parallel (red-solid line) directions to the axis of release.  Note that the scaling function of the direction in which the trapping frequency remains fixed (blue, dashed) in fact decreases during the expansion. The parameters used in this solution are $\omega_{||}=100$Hz, $\omega_{\perp}=10^3$Hz and a sudden release of the $||$ direction only.\label{scale}}
\end{center}
\end{figure}

In Fig.~\ref{scale}
we show the numerical solution to \eqref{scalingf} for cigar-like initial geometry after the potential in the ``$||$" direction is suddenly switched off while keeping the ``$\perp$" potentials constant. Notice that the tightly constrained ``$\perp$" dimension in fact pulls even tighter during the expansion in the ``$||$'' dimension as indicated by the decay of the scaling functions $\lambda_\perp$.  In this way one can understand how the volume of the cigar geometry, scaling as $(\lambda_{||}\lambda_\perp^2)^{-1}$, decays much more slowly than in an isotropically expanding BEC which scales as $\lambda_{||}^{-3}$.

\subsection{Validity of the hydrodynamic approximation}

\begin{figure}
\begin{center}
\includegraphics[scale=0.35]{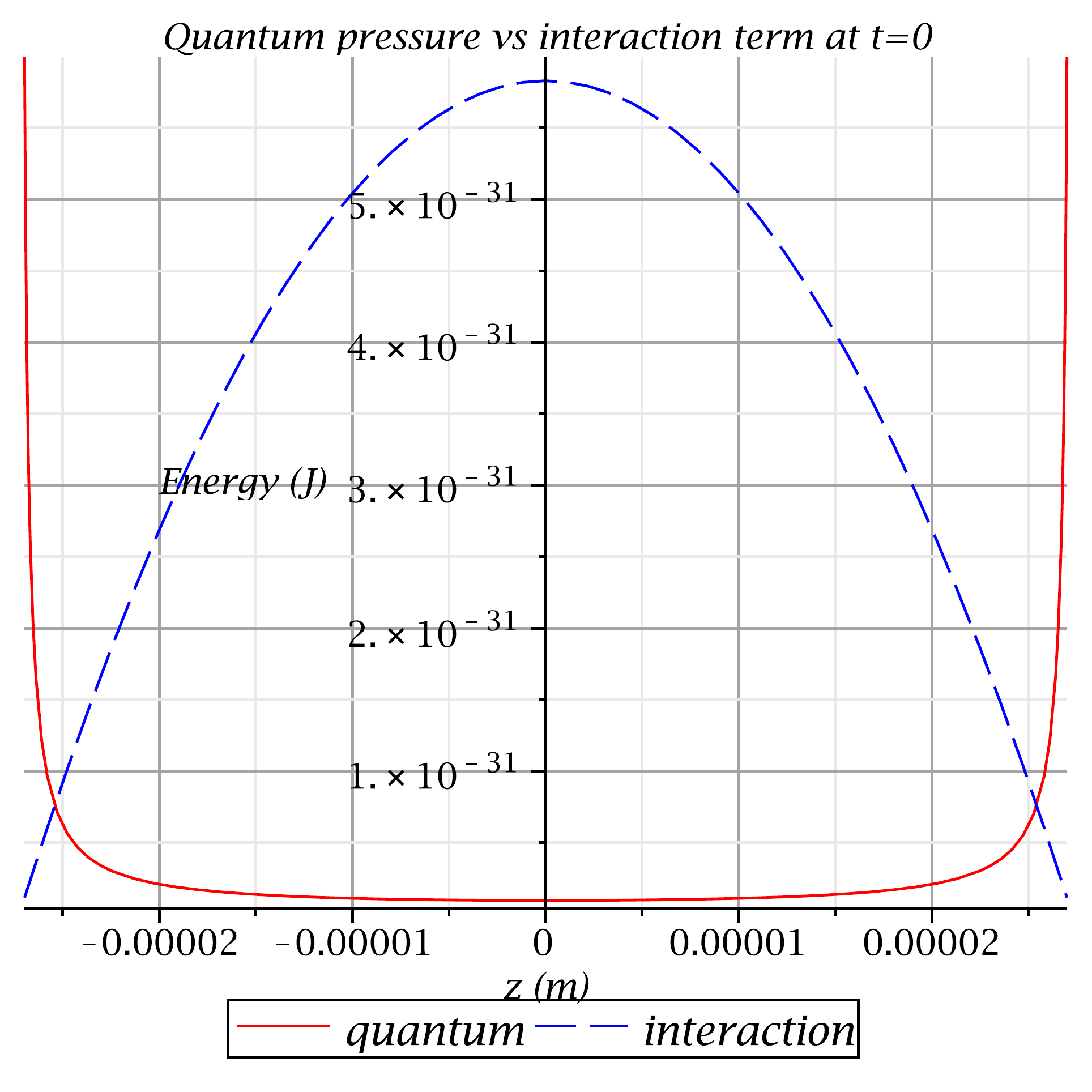}\caption{The quantum pressure and interaction energy expressed in Joules as a function of axial distance in meters. Note that the quantum pressure is certainly non-negligible near the boundary of the TF approximate ground state density wavefunction (at $z\simeq 2.7\times 10^{-5}$m).
 \label{quantum}}
\end{center}
\end{figure}

\begin{figure}
\begin{center}
\includegraphics[scale=0.35]{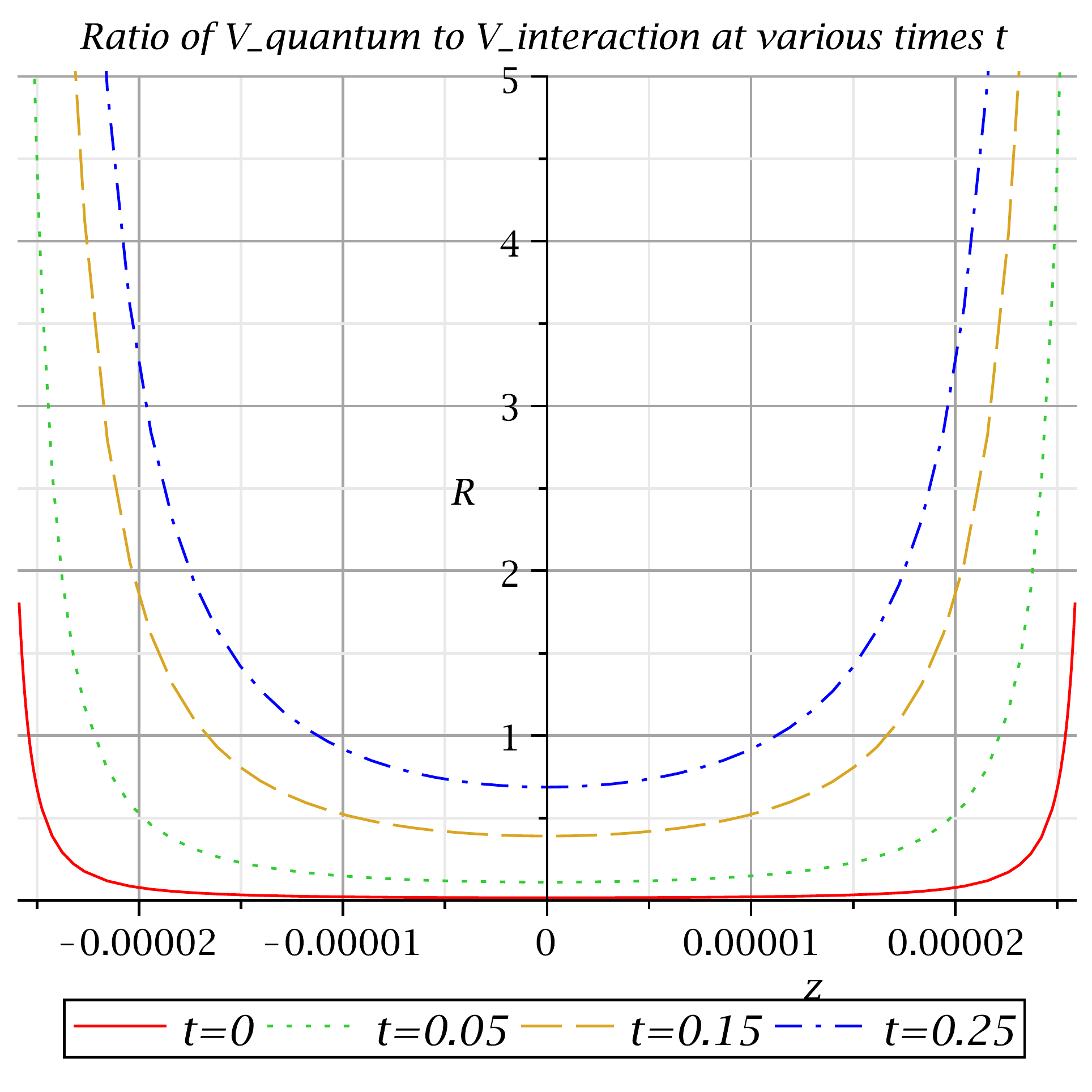}
\caption{The ratio $R=V_{\text{quantum}}/V_{\text{interaction}}$ as a function of co-moving axial position $\tilde{z}=z\lambda_{||}$ at various times $t$ after release of the trapping potential.  Note that the axial half-length of the condensate at time $t=0$ (which of course is the co-moving length for all times) is $L_z\simeq 2.7\times 10^{-5}$m. \label{quantumratio}}
\end{center}
\end{figure}

As already mentioned above, it is also important to keep track of the hydrodynamical approximation throughout the expansion. Since the quantum pressure scales differently with time than the density, it is of interest to understand whether the hydrodynamic approximation improves or degrades as the expansion proceeds and, if it loses accuracy, when the approximation breaks down.   In Fig.~\ref{quantum}
we display the quantum pressure $V_{\text{quantum}}$ and interaction $V_{\text{interaction}}$  ($=g\rho$) contributions to GP before the release of the axial trapping potential.  In Fig.~\ref{quantumratio}
we display the ratio $V_{\text{quantum}}/V_{\text{interaction}}$ as a function of co-moving axial position $\tilde{z}=z\lambda_{||}$ for four different times after the expansion.

Note that although there is always a region near the edge of the condensate where the ratio is greater then $1$, as time progresses the proportion of the condensate which remains within the hydrodynamical regime gets smaller until approximately $t=0.25$s when the quantum pressure becomes important globally in the condensate. The failure of the hydrodynamic approximation near the boundary of the condensate is expected in general since the TF approximation is known to break down on scales on the order of a healing length from the boundary where the density becomes exponentially small.  Indeed the true exact solution, being a smooth exponential decay to zero, differs from the TF approximate solution, given in \eqref{sol}, near the edge which sharply goes to zero there. Our analysis in the following will be concentrated on the more central part of the condensate where the TF approximation holds.

\section{Quantum Excitations}

We now intend to consider particle creation associated with the expansion of the condensate and its signature in the
correlation structure. Before doing so however, one might
wonder if such expansion is sufficiently rapid
to lead to any relevant particle production at all. Also, given that
we shall work in the analogue gravity framework, one would
also like to be sure that the ostensible excitations can be
meaningfully described as phonons rather than having to
deal with the complicated issues associated with non-linear
 dispersion. We shall discuss these issues first and then introduce
the correlation function that will be the subject of our studies.

\subsection{Timescales}

Let us now analyze and  compare the relevant inverse time scales involved in the problem of an expanding condensate to make an order of magnitude estimate of whether excitation effects can be expected to be observable.

Firstly there is an intrinsic infrared cutoff given by the size of the condensate. This length scale translates into a frequency $\omega_s$ using the (central maximum) sound speed $\omega_s=c_s/(L\lambda(t))$.

Secondly there exists an intrinsic ultraviolet cutoff associated with the healing length, the   healing frequency $\omega_h=2\pi c_s/\xi$ where $\xi=\hbar/(mc_s)$ is the healing length. The healing length, as already mentioned, is the length scale at which the quantum potential, neglected in the hydrodynamical approximation, becomes intrinsically comparable with the interaction energy (which is proportional to the density of the condensate and hence becomes less relevant at lower densities). This is also the scale at which the  Bogoliubov dispersion relation enters in the quadratic regime and hence the description in terms of non-interacting phonons breaks down.

Thirdly there is the characteristic frequency associated with the expansion itself, $\omega_e\approx\dot{\lambda}/\lambda$, which fixes the frequency of the most abundantly created excitations in an expanding condensate. This is generic to any particle creation effect by a time-varying external field (a ``dynamical Casimir effect") see e.g.~\cite{Barcelo:2003wu}.

\begin{figure}[h]
\begin{center}
\includegraphics[scale=0.4]{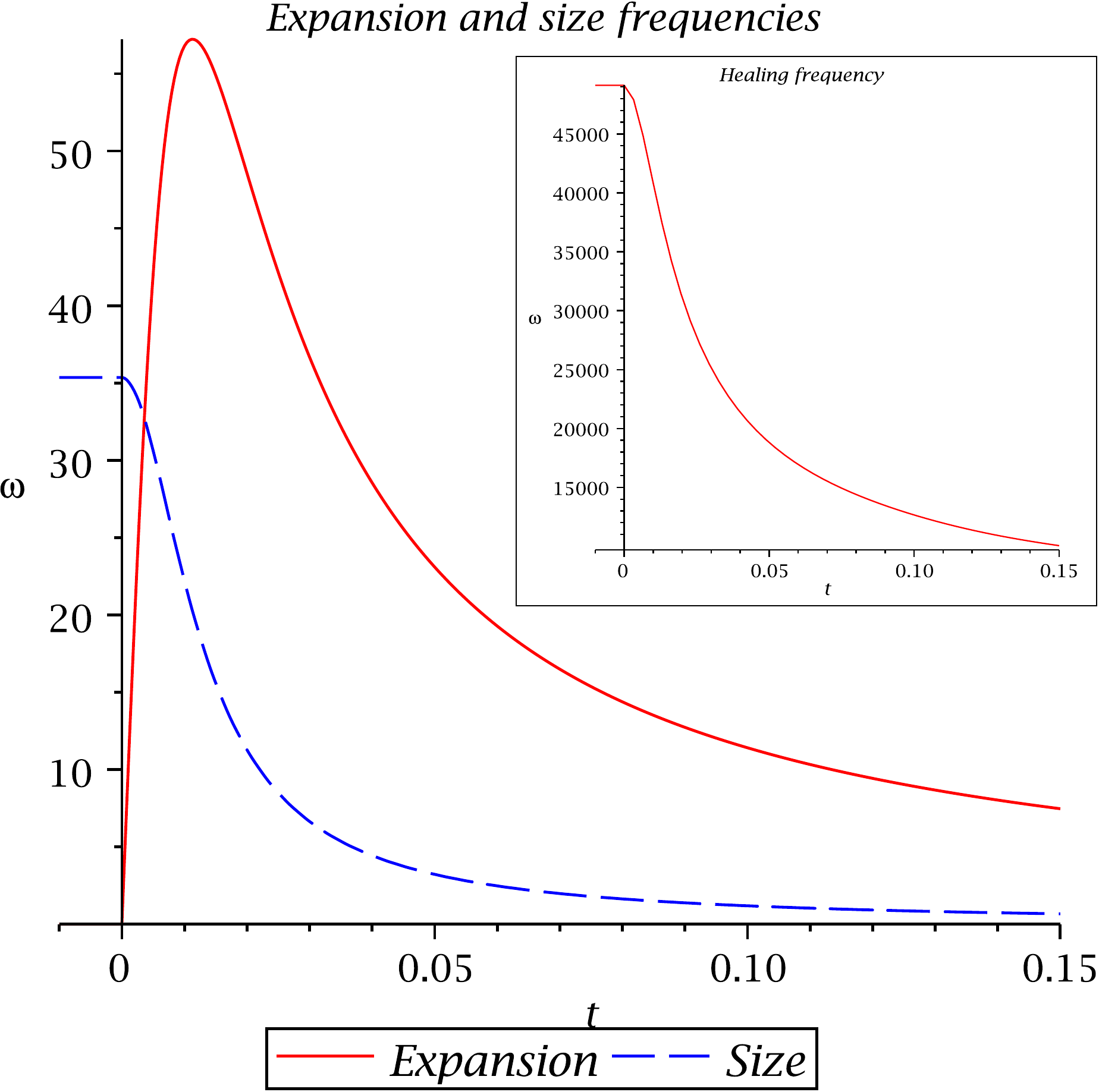}
\caption{A comparison of the relative magnitudes of $\omega_e$ and $\omega_s$ (red-solid and blue-dashed respectively) and with the healing frequency (inset plot) in $s^{-1}$ as a functions of time (in $s$). Note that, for the majority of the duration of the expansion, $\omega_s$ is much smaller than the expansion frequency $\omega_e$ which is much smaller than the healing frequency. Hence the bulk of the particle production spectrum lies within the low frequency cutoff provided by the total size of the condensate and the high frequency cutoff provided by the healing length.  \label{typical} }
\end{center}
\end{figure}

It is of interest to compare these three frequencies in order to test the validity of our approximations and to estimate if the expected excitations are created within the natural ultraviolet and infrared cutoffs. In Fig.~\ref{typical} we compare these three inverse time scales using the accurate parameters listed above.

We see that indeed there exists a regime in which the typical excitation frequency lies entirely within the natural ultraviolet and infrared cutoff frequencies. Quantitatively this regime begins after about $0.005$s.  Let us also note the vastly higher healing frequency representing the UV cutoff for a phononic treatment. This is shown in the smaller subfigure in Fig.~\ref{typical} for comparison.

Therefore we conclude that there is a viable regime within which the excitations have a phononic (linear) dispersion relation and wavelengths much shorter than the size of the BEC.

\subsection{Wightman functions}

The Wightman function $G(x,y):=\langle \phi(x) \phi(y)\rangle$ for a quantum field $\phi$ is a measure of the correlation between two separate points $x$ and $y$ in spacetime.  If the field $\phi$ were a truly random variable the correlation would be zero everywhere.  In general the equal-time Wightman function for a massless scalar field on a flat 3+1 dimensional FRW spacetime with scale factor $a$ (which includes Minkowski space as the $a=1$ special case)\footnote{Recall that a FRW spacetime is described by the metric $ds^2=a^2(\eta)\left(-d\eta^2+d\mbf{x}^2\right)$ here written in co-moving coordinates $\mbf{x}$ and conformal time $\eta$ where $a(\eta)$ is a scalar function of time.}  only depends on the magnitude of the co-moving spatial difference $x=||\mbf{x}-\mbf{y}||$ and (possibly) on time.  This homogeneous function thus is really only a function of these two variables
\begin{align}
G(\eta, \mbf{x}, \mbf{y})&=\int \frac{d^3kd^3k'}{(2\pi)^{3}}\langle 0|\phi_k\phi_{k'}|0\rangle\text{e}^{i(\mbf{k}\cdot\mbf{x}+\mbf{k}'\cdot\mbf{y})} \nonumber \\
&=\int_0^\infty \frac{dk}{2\pi^2}\frac{k}{x}\,\text{sin}\,(kx)\;u_k(\eta)u_k^*(\eta)\nonumber  \\
 &=G(\eta, x) \label{Wight}
\end{align}
where $k=\sqrt{\mbf{k}\cdot\mbf{k}}$, $u_k$ are the mode-function
solutions to $u_k''+(k^2-a''/a)u_k=0$  and
$\phi_k=(b_ku_k+b^{\dag}_{-k}u_k^*)/(\sqrt{2}a(\eta))$,  prime
denoting the conformal time derivative $\partial_\eta$.   In
Minkowski spacetime the integral is computed exactly as
$G(\eta,x)=\hbar/(4\pi^2x^2)$ for non-zero $x$ while on the
light-cone the Wightman function also possesses a singular imaginary
part \cite{fulling89}.  In 1+1 dimensions the equal time Wightman function
becomes logarithmic and time independent $G(x,y)=-\hbar/4\pi\;\text{ln}\,(x-y)^2$
and, owing to the conformal invariance in 1+1 dimensions for the
massless scalar field and the conformal flatness of FRW spacetimes,
also takes this form in an arbitrary 1+1 FRW background in terms of
co-moving coordinates and the conformal time variable \cite{Birrell:1982ix}. 

\section{Correlations in a isotropically expanding BEC}


The main purpose of this section is to gain an intuitive understanding of correlations due to particle production and provide a useful toolbox for the more challenging investigation of the anisotropic scaling solution. In order to do so we shall study first the ideal case of  a homogeneous 3+1D isotropically expanding BEC. In this case, the analogue metric is that of a simple FRW spacetime geometry. We shall choose a scale factor that increases from an initial constant value $a_i$ to a final constant value $a_f$ smoothly as in this case the structure of correlations can be solved for analytically. Our interest in these exactly soluble models derives from the fact that in these systems one can see the appearance in the Wightman function of the characteristic features expected from the creation of particles in a transparent way. Finally, we shall also discuss how to simply interpret such models in terms of a BEC with a varying scattering length.

\subsection{Exactly Soluble Expanding Spacetime \label{threeplusone}}

Written in terms of co-moving coordinates $\tilde{x}_i=x_i/\lambda_i(t)$ the metric \eqref{back} for the scaling solution BEC reads
\be
ds^2=\sqrt{\frac{\rho(t,\mbf{x})}{gm}}\left(-c_s^2(t,\mbf{x})\,dt^2+\lambda_i^2(t)\, d\tilde{x}^i{}^2\right) \label{analogue}
\ee
from which a number of standard representations for a FRW background are easily obtained.  The cosmological assumption of homogeneity is satisfied for a BEC in the central region where the density is almost constant.  Working in such a central region, $\rho$ and $c_s$ become functions only of $t$ and the metric can be written in standard FRW form as
\be
ds^2=-d\tau^2+a^2(\tau)d\tilde{\mbf{x}}^2, \quad \tau(t)=\int^t dt' \frac{\rho^{3/4}(t')g^{1/4}} {m^{3/4}} \label{FRW}
\ee
with the scale factor defined by
\be
a^2(\tau)=\sqrt{\frac{\rho(\tau)}{gm}}\lambda^2(\tau)\propto\sqrt{\lambda(\tau)}
\ee
where we have used the scaling solution $\rho(\eta)=\rho_0/\lambda^3(\eta)$ for the density.

In what follows however, we will work with the form for the metric given by
\be
ds^2=-a^6(\eta)\,d\eta^2+a^2(\eta)\,d\mbf{x}^2 \label{background}
\ee
which can be obtained form the form \eqref{FRW} by the coordinate transformation
\be
d\eta=\left(\frac{gm}{\rho_0}\right)^{3/2}\frac{1}{\lambda^{3/2}(\tau)}d\tau.
\ee
The functional relationship between the scale factor $a$ the scaling function $\lambda$ is unchanged in this representation as $a^2(\eta)\propto\sqrt{\lambda(\eta)}$.

Furthermore we will consider the specific form for the scale factor
\be
a^4(\eta)=\frac{a_i^4+a_f^4}{2}+\frac{a_f^4-a_i^4}{2}\,\text{tanh}\left(\frac{\eta}{\eta_0}\right) \label{scalef}
\ee
where $\eta_0$, $a_i$ and $a_f$ are constant numerical parameters.
Such a metric describes an isotropic homogeneous spacetime with flat
spatial sections which expands by a finite amount over a finite time
period.  This spacetime has been studied previously in the
literature for example in \cite{Barcelo:2003wu} where it was shown, to arise as the analogue geometry experienced by
phase perturbations in a BEC with time varying interaction strength
$g$. In that article the particle content of the initial Minkowski
vacuum state after expansion is studied in detail and the
observability of the created particles discussed.  This spacetime is
also studied in \cite{PhysRevLett.99.201301} where it is shown to
arise as the analogue geometry associated with a ring of trapped
ions. A very similar geometry is studied in \cite{Carusotto:2009re}
where instead of $\tanh$, the authors employ an error function\footnote{Recall that the error function is defined as Erf$(t)\propto\int^t
\text{exp}(-x^2)\,dx$ and smoothly interpolates between constant values.} temporal profile for the variable scattering length
and the correlation structure is studied semi-analytically in the
so-called ``sudden limit" (which we will discuss later), and
numerically in the general case, with the full Bogoliubov spectrum. In \cite{PhysRevA.79.033601} the author studies the controlled release of a condensate such that the trapping frequency varies exponentially in time and the associated particle production in terms of the number of particles produced. The separate case of an exponential variation of a time dependent coupling constant $g(t)$ is also studied as well as the mixed analysis where both the trapping frequency and coupling constant are varied again in terms of the number of particles produced.

The field equation for a massless minimally coupled scalar field $\phi$ propagating on the background geometry \eqref{background} decouples into the independent mode equations
\be
\partial^2_\eta\phi_k+a^4k^2\phi_k=0 \label{fieldeq}
\ee
where $\phi_k=(2\pi)^{3/2}\int d^3k\,\text{e}^{-i\mbf{k}\cdot\mbf{x}}\phi$ is the kth Fourier component of the field $\phi$.   For $a(\eta)$ as \eqref{scalef} the field equation \eqref{fieldeq} is solved by a rather complicated product of elementary and hypergeometric functions \cite{Barcelo:2003wu}.  Conveniently for us, it is not necessary to work explicitly with these cumbersome functions: For each co-moving momentum $k$ the physics of cosmological particle production is contained only in inner products (which we label as $\alpha_k$ and $\beta_k$) between two particular solutions $\phi_k^{\text{in}}$ and $\phi_k^{\text{out}}$ to \eqref{fieldeq}, solutions which converge respectively in the past ($\eta\rightarrow -\infty$) and future ($\eta\rightarrow\infty$) to plane waves\footnote{Specifically $\alpha_k$ is the innerproduct between positive frequency plane waves whereas $\beta_k$ is the inner product between one positive and one negative frequency plane wave} (see for example the standard monograph \cite{Birrell:1982ix}).

\begin{figure}
\begin{center}
\includegraphics[scale=0.3]{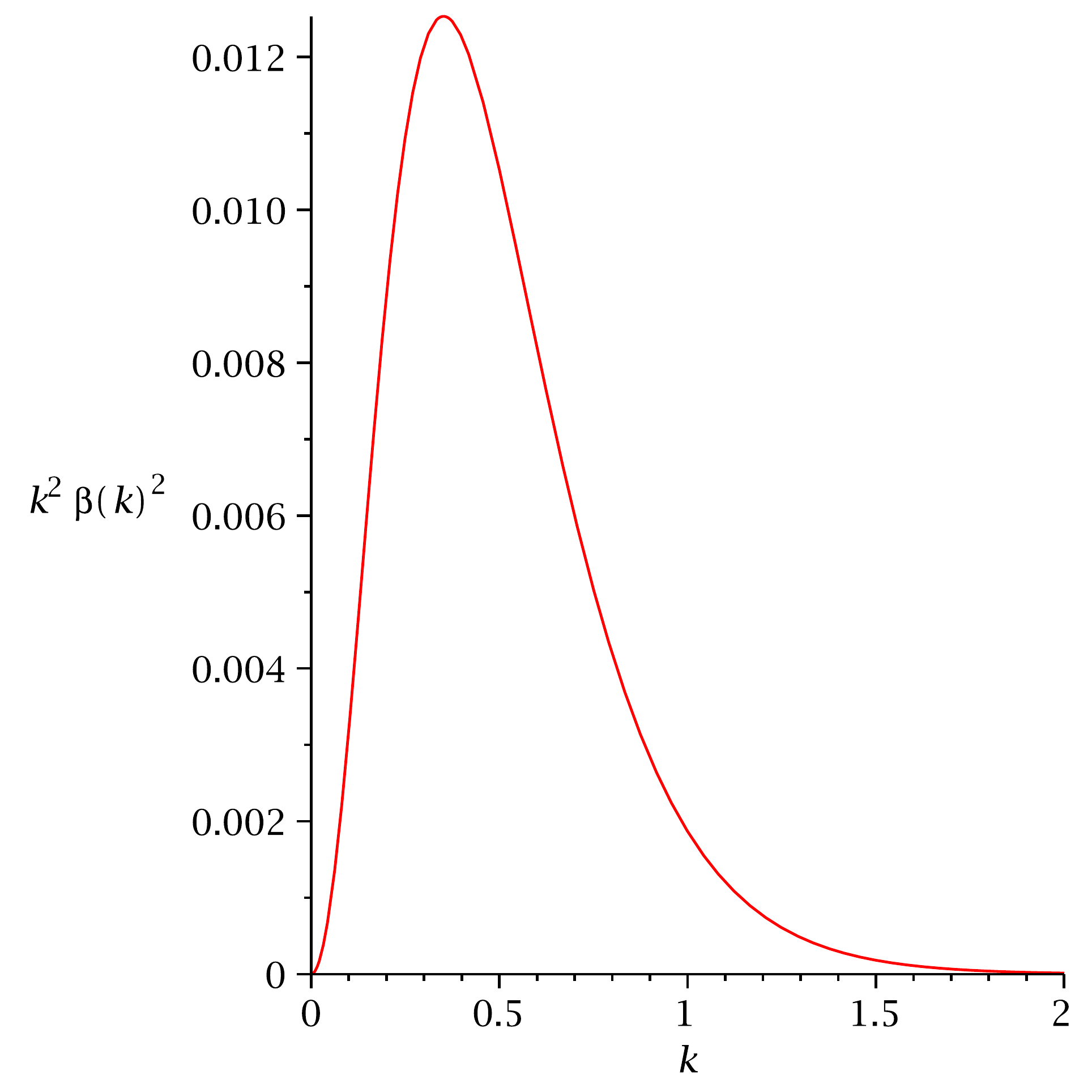}\caption{The characteristic shape of the spectrum of produced particles in an expanding spacetime described by \eqref{background}, as a function of the modulus $k=\sqrt{\mbf{k}\cdot\mbf{k}}$. Recall that the spectrum is proportional to the volume of space $V/2\pi$ as $N_k=V k^2\beta_k^2/2\pi$ so that what we plot here is in fact particle number density. {The peak wave number is fixed by the typical timescale of the expansion $k_{peak}\propto 1/\eta_0$}. \label{Spec}  }
\end{center}
\end{figure}

Physically, these two particular solutions represent a description of two different vacuum states, $|\text{in}\rangle$ and $|\text{out}\rangle$, in the two asymptotic regimes where the function $a(\eta)$ becomes constant and the definition of particle state is unambiguous. For example, the quantum state $|\text{in}\rangle$ described by the choice $\phi_k^{\text{in}}$ (the vacuum state for $\eta\rightarrow -\infty$) is no longer a vacuum (zero particle) state as $\eta\rightarrow\infty$  since that state, $|\text{out}\rangle$, is described by the mode functions $\phi_k^{\text{out}}$ and in general one has
\be
\phi_k^{\text{in}}=\alpha_k\phi_k^{\text{out}}+\beta_k\phi_{k}^{\text{out}}{}^*
\ee
with $\beta_k\neq0$.  The spectrum of particle content of the state $|\text{in}\rangle$ at late times in terms of the late time particle states is given by the modulus $|\beta_k|^2$ which has the characteristic bell shape shown in Fig. \ref{Spec} here shown as a function of the isotropic modulus $k=\sqrt{\mbf{k}\cdot\mbf{k}}$ appropriate for the isotropic mode equations \eqref{fieldeq} and hence isotropic coefficients $\alpha_k$ and $\beta_k$.

The inner products $\alpha_k$ and $\beta_k$ (known as Bogoliubov coefficients) are given exactly in this model by the simple expressions
\begin{align}
\alpha_k &=\frac{2\sqrt{AB}}{A+B}\frac{\Gamma(-iA)\Gamma(-iB)}{\Gamma^2\left(-i(A+B)/2\right)}\nonumber \\
&\label{eq:ab}\\
\beta_k &=\frac{-2\sqrt{AB}}{B-A}\frac{\Gamma(-iA)\Gamma(iB)}{\Gamma^2\left(i(B-A)/2\right)}\nonumber
\end{align}
where $\Gamma$ is the Euler function and the dimensionless variables $A$ and $B$ are
\be
A=k\eta_0a_i^2, \quad B=k\eta_0a_f^2.
\ee
They are constrained to satisfy $|\alpha_k|^2-|\beta_k|^2=1$ by the constancy of the Klein Gordon norm of the (unique) solution $\phi$ written in the two different mode function bases.


Since the mode-functions are simple linear combinations of plane waves long after the expansion has taken place, the equal time Wightman function there can be written down in a relatively simple way in terms of $\alpha_k$ and $\beta_k$ alone as
\be
G(\eta,x)=\frac{4\pi}{2a_f^2}\int^\infty_0 dk\left[\frac{\text{sin} \,kx}{x}\left(1+2|\beta_k|^2\right)+I+I^*\right] \label{int}
\ee
where
\be
I=\frac{\text{sin}\;kx}{x}\alpha_k\beta_k^*\text{e}^{-2ika_f^2\eta}.\label{eq:I}
\ee
Note again that, given spatial homogeneity and isotropy of the system, here $G$ is a function only of the magnitude of the spatial separation.
Of course, such an expression cannot be considered exact for a realistic finite-sized BEC but will break down nearby the boundaries due to finite size effects. Furthermore, in the case of a finite volume ``cigar"-shaped BEC and anisotropic expansion the expression \eqref{int} will have to be replaced by a more complicated formula as we shall see in sect.~\ref{sec:elongated}.

\begin{figure}
 \begin{center}
  \includegraphics[scale=0.4]{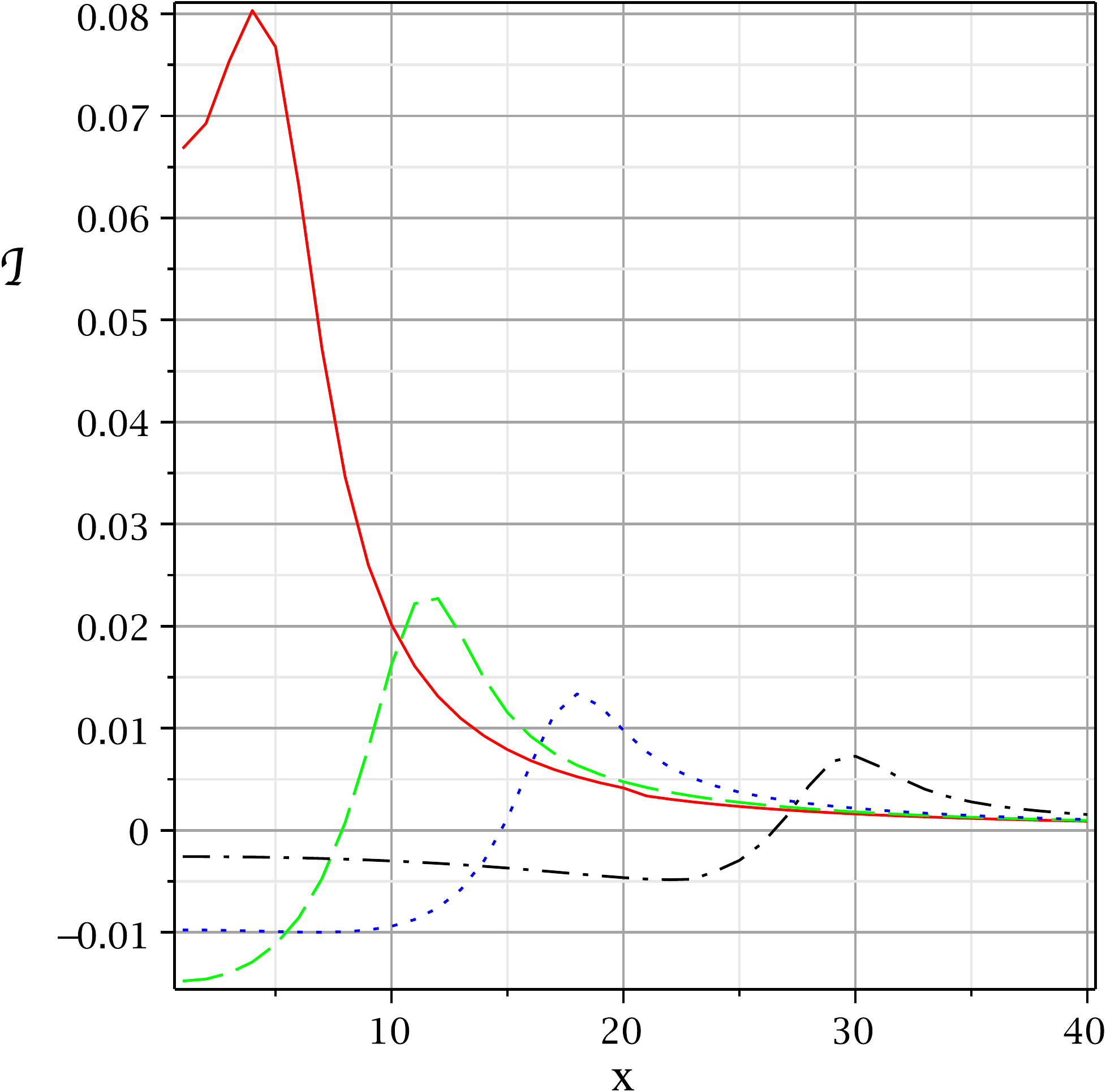}\caption{The exact numerical integration of (minus) the entanglement contribution to $G$ as a function of point separation $x$ for four successive times $\eta=0$ (red-solid), $\eta=1$ (green-dashed), $\eta=2$ (blue-dotted) and $\eta=4$ (black-dot-dashed) for the choice of parameters $a_i^2=1$, $a_f^2=\pi$ and $\eta_0=1$. \label{numerical}}
 \end{center}
\end{figure}

The expression \eqref{int} is the sum of three physically distinct terms.  The $1$ term represents the background zero point fluctuations, the $|\beta_k|^2$ term an enhancement of the background correlations due to particle production
and the terms labelled $I$ and $I^*$ represent additional non-trivial structure propagating on top of the (enhanced) Minkowski-like correlations due to the entanglement between the created pairs of particles.  The $I+I^*$ term can be integrated exactly as an unwieldy infinite sum of functions which we do not write down here.  In Fig. \ref{numerical}
we plot the result of numerically integrating the contribution from $I+I^*$
\be
\mathcal{I}=\frac{4\pi}{2a_f^2}\int^\infty_0 dk \left(I+I^*\right)
\ee
in $G$ for four successive times as a function of separation $x=||\mbf{x}-\mbf{x}'||$. We see that going backwards in time a bump of correlations merges into the correlation singularity\footnote{Since we always work under the assumption that the field theory is an effective field theory valid only above a small length scale, for example the healing length in the case of a BEC, the coincidence singularity does not exist in practice.  In the BEC example, higher order contributions to the dispersion relation become non-negligible and in fact dominate at very short length scales resolving the singularity.} at the origin.  The (coordinate) speed of the bump is determined by the final value of the scale factor $a_f^2$ which in the displayed plot was taken to be $\pi$. We note that the bump moves with a velocity of close to $7.5$ units per second which is approximately the speed at which two particles moving at speed $\pi$ units per second would be receding from one another.   Note that the coordinate speed is not constrained to be the speed of sound $c_s$, the coordinate speed in lab variables.


In the next section we show that the essential structure of these correlations are present already in the so-called adiabatic approximation to the integrand $I+I^*$.

\subsubsection{Adiabatic and sudden approximations}

The adiabatic approximation for a mode $k$ is accurate when $k\eta_0\gg1$.  It is a good approximation for high wave number, or short wavelength modes. These are the modes which feel the expansion as a ``slow'' process. In the adiabatic limit one can show, from \eqref{eq:ab} and \eqref{eq:I}, that
\be
I+I^*\simeq\frac{2}{x}\;{\sin}\;(kx)\;\sin\left[k\left(\Phi-2a_f^2\eta\right)\right]\text{e}^{-\pi a_i^2 k\eta_0}
\ee
where the phase is given by
\be
\Phi=\ln\;\left(\frac{a_i^2+a_f^2}{a_f^2-a_i^2}\right)a_i^2\eta_0+\ln\;\left(\frac{a_f^4-a_i^4}{4a_f^4}\right)a_f^2\eta_0.
\ee

\begin{figure}
 \begin{center}
  \includegraphics[scale=0.35]{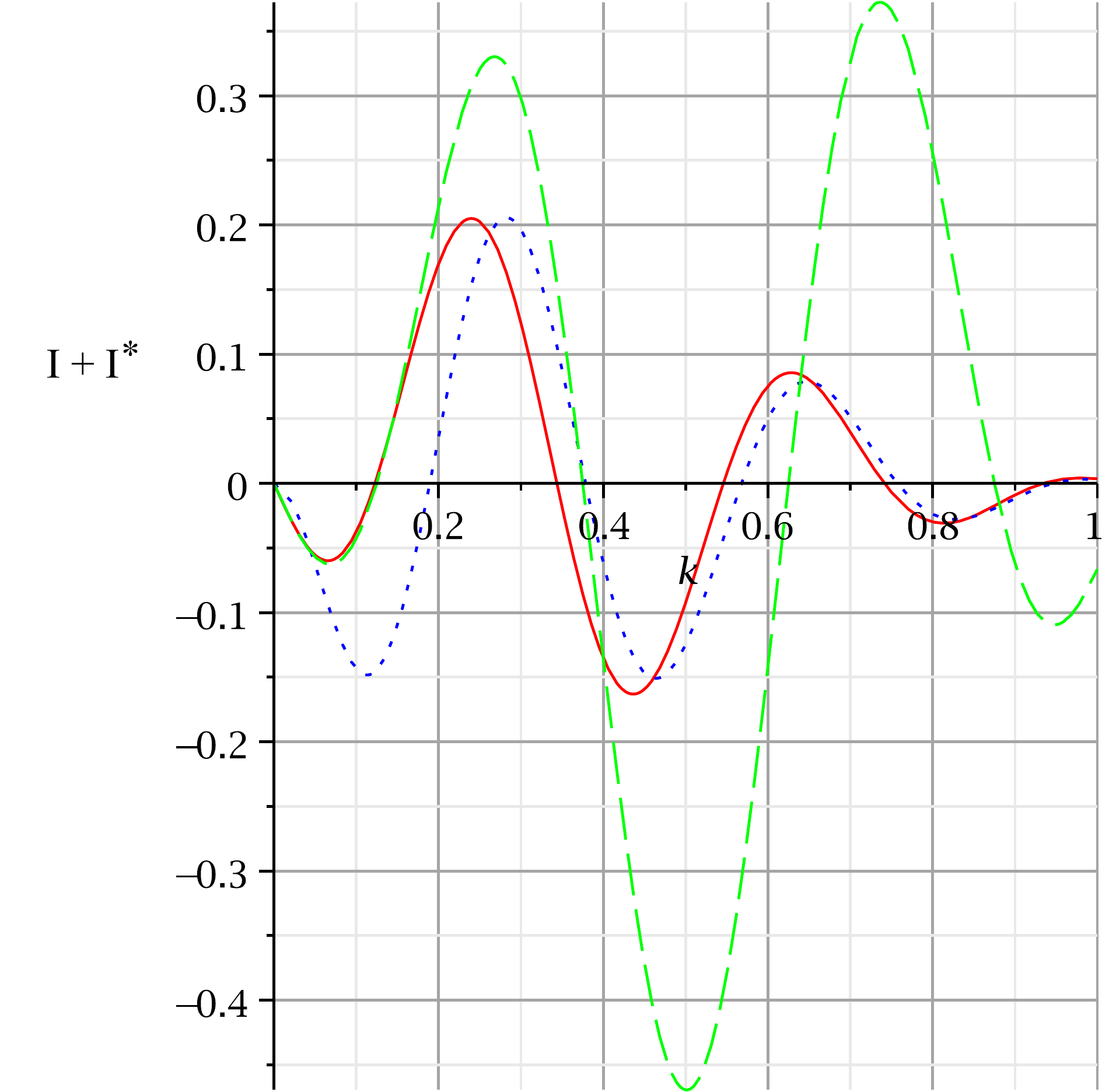}\caption{A comparison of the exact expression (in red-solid) with both the adiabatic approximation (in blue-dots) and the sudden approximation (in green-dashes) near the turn over region ($k\eta_0\approx 0.3$) where neither approximation is very accurate.  We use the representative values of  $t=2$ and $x=3$. Note that the sudden approximation is most accurate for small $k$ and the adiabatic for large $k$. Also note that the adiabatic approximation remains relatively accurate (at least in mimicking the qualitative behavior of the exact function) all the way down to $k=0$ where the approximation in principle should be poor.
\label{fullcomp}}
\end{center}
\end{figure}

On the other hand, the sudden approximation is accurate for  a given $k$ whenever $k\eta_0\ll1$. Hence it should be a good approximation to the exact result for low wavenumbers.  In this sudden limit one can similarly show (again from \eqref{eq:ab} and \eqref{eq:I}) that
\be
I+I^*\simeq -\frac{1}{2}\frac{a_f^4-a_i^4}{a_f^2a_i^2}\frac{\sin\;kx}{x}\cos(2ka_f^2\eta)\, \label{sudd}
\ee
which represents an unbounded contribution to $G$ if assumed accurate over the entire positive $k$ axis (as would be the case in the formal limit $\eta_0\rightarrow 0$).  In Fig. \ref{fullcomp} we compare the exact, adiabatic approximate and sudden approximate forms of $I+I^*$ for a representative choice of $x$ and $\eta$.
We note that the adiabatic approximation is accurate almost uniformly over all $k$, even close to the origin, where it is superseded by the sudden approximation while still capturing the qualitative behavior of the $I+I^*$ term.

\begin{figure}
\begin{center}
 \includegraphics[scale=0.35]{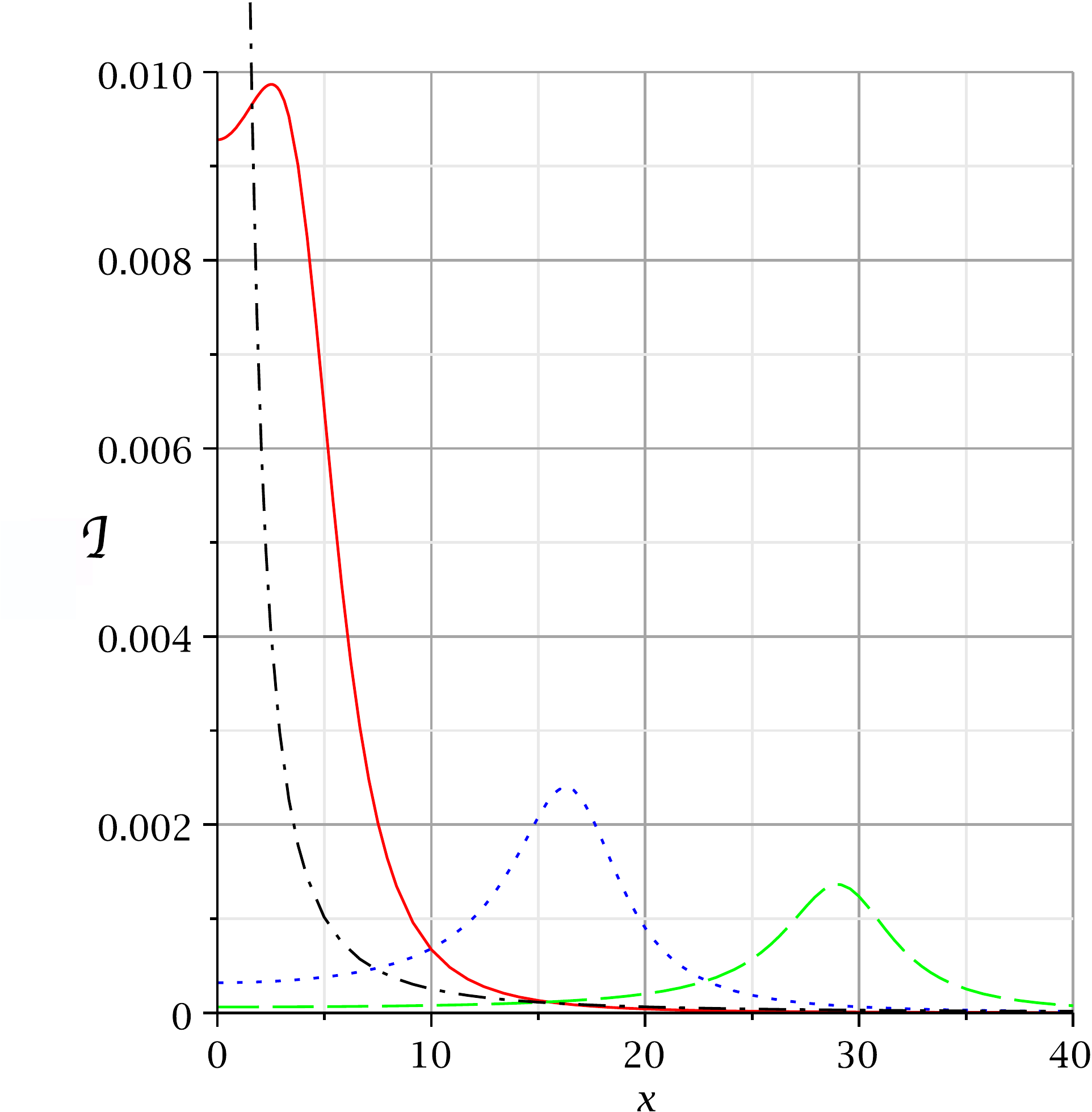}
\caption{The integrated $I+I^*$ contribution to the Wightman function in the adiabatic approximation at three successive times $\eta=0$ (red-solid),  $\eta=2$ (blue-dots) and $\eta=4$ (green-dashes). Note that since the expression \eqref{int} is only valid after the expansion has taken place (the midpoint of which occurs around $\eta=0$) the contribution for $t<0$ are not shown. To make these plots and for comparison with the exact result again we have used the values $a_i^2=1$, $a_f^2=\pi$, and $\eta_0=1$. Also plotted (in black- dash-dot) is the standard flat spacetime correlations $(4\pi^2 x^2)^{-1}$. \label{bump}}
\end{center}
\end{figure}

Integrating the adiabatic approximation $I+I^*$ over $k$ we observe a ``bump" of correlations traveling out from the origin and decaying as displayed in Fig. \ref{bump}. This is in close agreement with the exact result as shown in Fig.~\ref{numerical}.
We note that the only feature of the exact result not captured by the adiabatic approximation is the negative ``tail" to the bump as it propagates outwards from the origin.

\begin{figure}[htb]
\begin{center}$
\begin{array}{cc}
\includegraphics[scale=0.20]{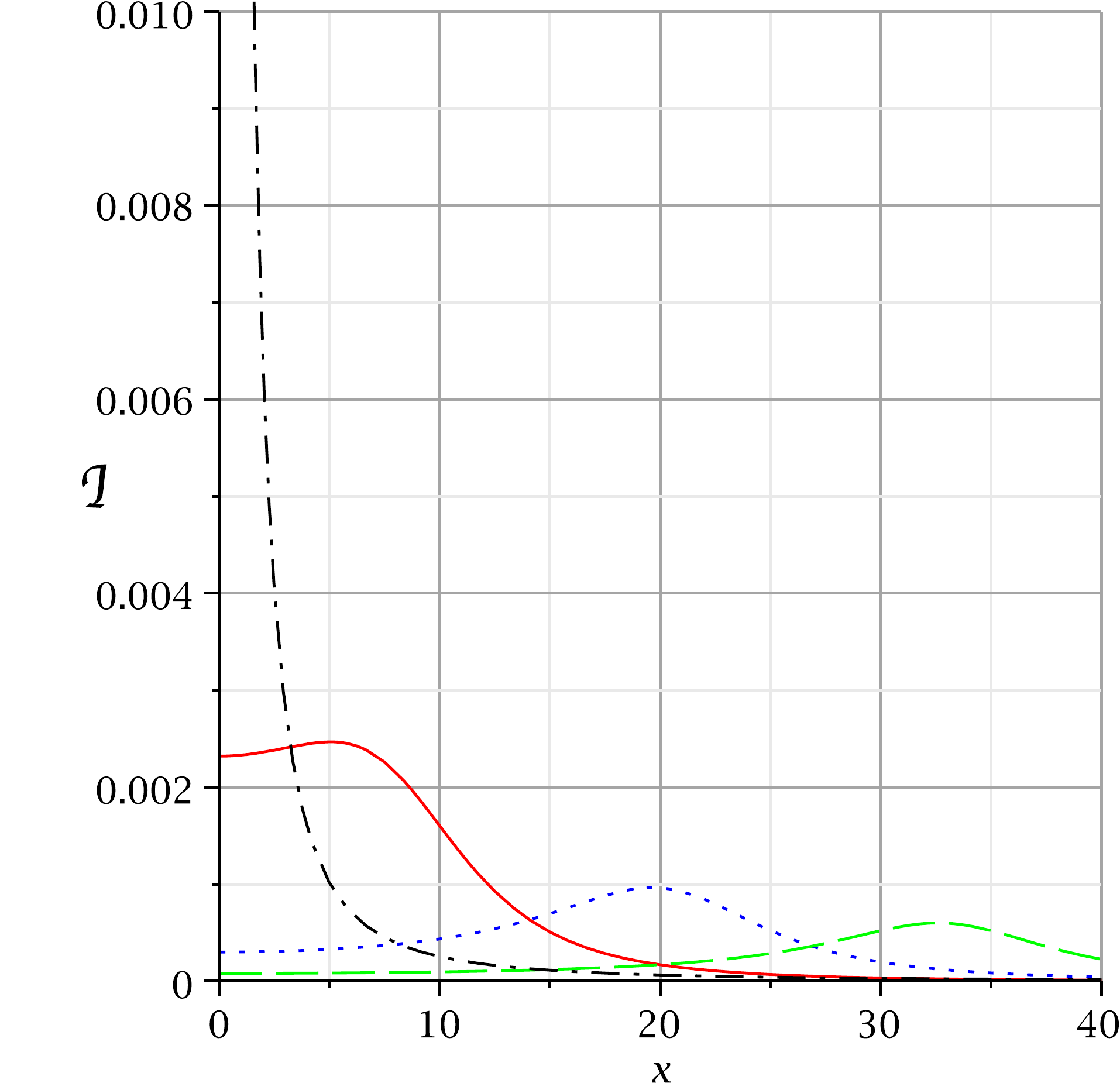} &
\includegraphics[scale=0.20]{compare} \\
\includegraphics[scale=0.20]{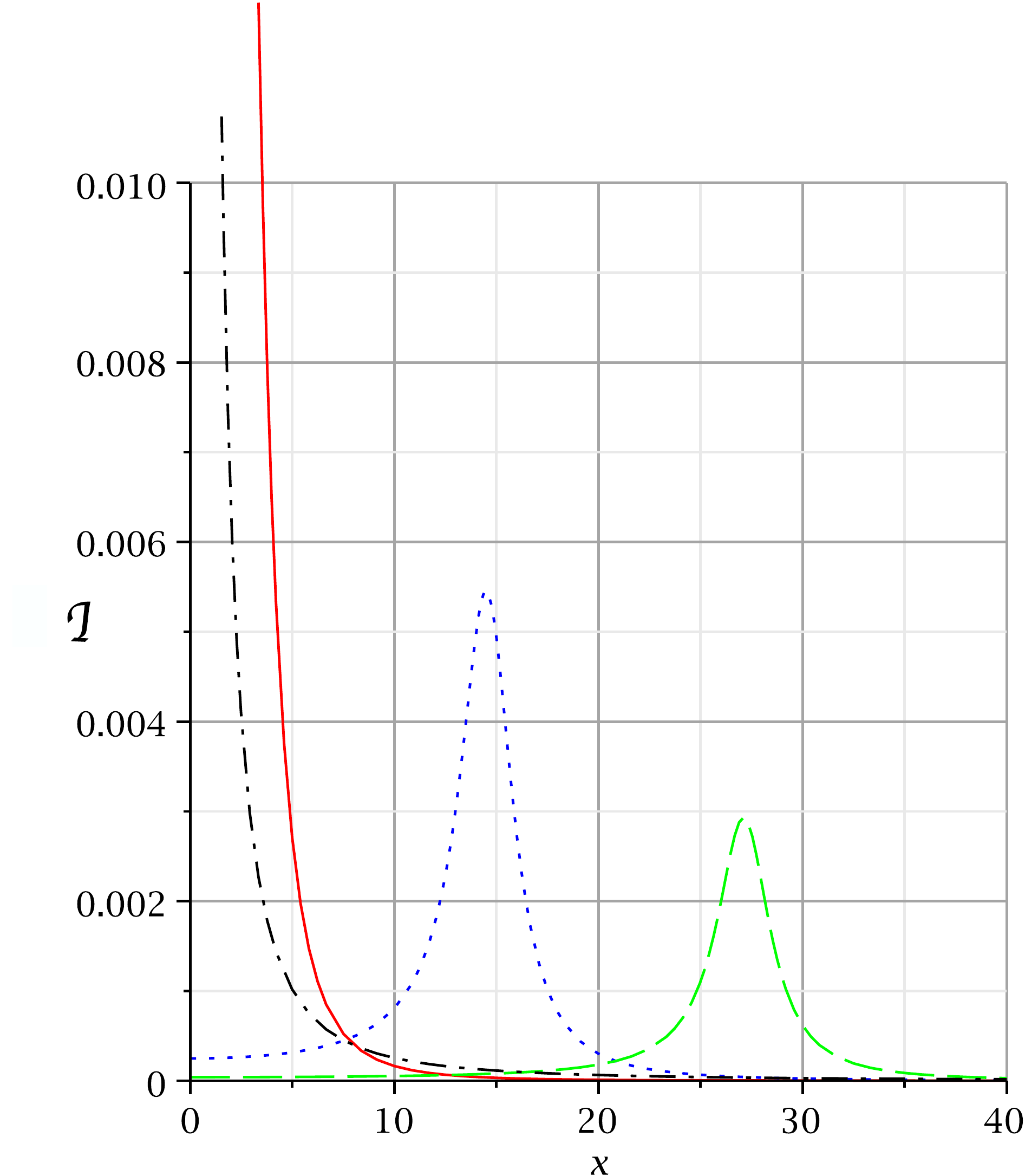} &
\includegraphics[scale=0.2]{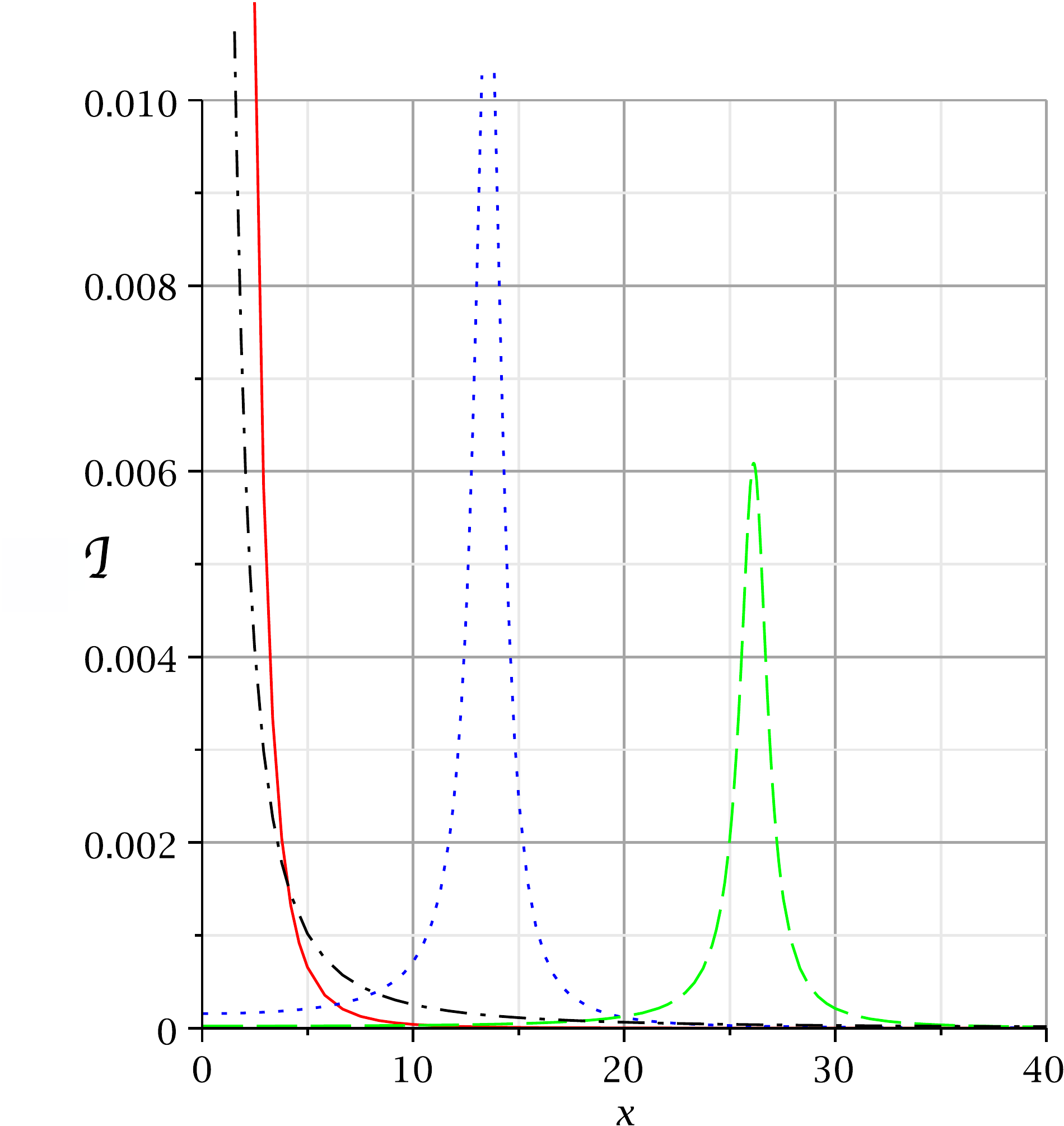}
\end{array}$
\caption{Comparison of the bump structure of correlations for three different values of the parameter $\eta_0=2$, $1$, $1/2$ and $1/4$ (from top left to bottom right) at three different times $\eta=0$ (red-solid),  $\eta=2$ (blue-dots) and $\eta=4$ (green-dashes). Also plotted (in black- dash-dot) is the standard flat spacetime correlations for comparison.}
\label{bumpcompare}.
\end{center}
\end{figure}

The role of the parameter $\eta_0$ is visible in Fig. \ref{bumpcompare} where we again show the result of integrating the adiabatic approximation\footnote{For computational reasons we restrict our attention to the adiabatic approximation in this section.}  to $I+I^*$ for the same values $a_i^2=1$ amid $a_f^2=\pi$ and times $\eta=0$, $2$ , $4$.
One sees that as the rate of expansion increases (as $\eta_0$ becomes small) the shape of the bump becomes sharply peaked and that the peak occurs slightly closer to the origin.  This is consistent with our intuitive understanding of the bump as arising from propagating particles created during the expansion.

For small $\eta_0$ the expansion and hence the particle production takes place over a short time interval.  Assume for the purposes of illustration that all the particle production occurs at a one point in space, say at $X$. (Particle production occurs at \emph{every} point but this simplification serves to capture the relevant physical mechanism) Then, two observers at a fixed spatial separation symmetric to $X$ will simultaneously observe a burst of particles passing by. For large $\eta_0$ the particle production is spread out over a longer time scale since the scale factor $a$ is changing significantly over a longer period and hence the observers will see a more diffuse burst of particles.   In this way it is not surprising that attempting to integrate the sudden approximate solution over all $k$ (that is, in the formal $\eta_0\rightarrow 0$ limit) one finds a propagating singularity at approximately $x=2a_f^2\eta$ as can be inferred from the structure of \eqref{sudd}.

\subsubsection{Varying scattering length interpretation}

As mentioned at the beginning of this section, the metric \eqref{background} can be used to describe phase perturbations in a BEC with a time dependent scattering length $\bar{a}(t)$ (and hence coupling parameter $g(t)$ and sound speed $c_s$) as discussed in the article \cite{Barcelo:2003wu}.   In such varying scattering length models one works in a homogeneous approximation where the background condensate density $\rho$ is assumed constant and the flow velocity $\mbf{v}$ everywhere zero. The FRW scale factor (which we temporarily write as $a_s$ for clarity) is related to the scattering length $\bar{a}$ by
\begin{align}
a_s(\eta)=\left(\frac{\rho}{4\pi \hbar^2}\right)^{1/4}\frac{1}{\bar{a}^{1/4}(\eta)}.
\end{align}
Hence a decrease in the scattering length corresponds to an expansion of the FRW spacetime and vice versa.  In \cite{Barcelo:2003wu} the authors use this form for the metric and also the tanh scale factor profile \eqref{scalef} described above.

We interpret our result in this context as a propagating bump of correlations when the scattering length is varied according to
\be
\bar{a}(\eta)=\frac{\rho}{4\pi\hbar^2}\left(\frac{a_i^4+a_f^4}{2}+\frac{a_f^4-a_i^4}{2}\;\text{tanh}\left(\frac{\eta}{\tau_0}\right)\right)^{-1}.
\ee
Note that this time dependence is not determined by any internal dynamics of the BEC but instead is a free experimental choice.  This is in contrast with the expanding BEC case where the time dependent parameter, the scaling function $\lambda$, is determined by the internal dynamics of the BEC.

\section{Axial correlations in an expanding elongated condensate}
\label{sec:elongated}


We shall now consider the experimentally relevant case of an anisotropic harmonically trapped BEC, where the trap is released only along the $z$ dimension.

We write $\lambda(t):=\lambda_z(t)$ and $\lambda_\perp(t):=\lambda_x(t)=\lambda_y(t)$. The simple form and cylindrical symmetry of the scaling solution in this case allows us to separate the variables in the background velocity as $c_s(t,z,r)=\tilde{c}(\tilde{z},\tilde{r})/\sqrt{\lambda(t)\lambda_\perp^2(t)}$ where $\tilde{c}=c_s\sqrt{1-\tilde{\omega}_z\tilde{z}^2-\tilde{\omega_r}\tilde{r}^2}$ and $\tilde{\omega}_i$ was defined in section \ref{scaling}.

Let us now introduce new coordinates through the exact differential expressions
\be
dT=\frac{dt}{\lambda^{3/2}\lambda_\perp}, \quad dZ=\frac{d\tilde{z}}{\tilde{c}(\tilde{z},\tilde{r})}+f\,d\tilde{r}\label{exact}
\ee
with $f(\tilde{r},\tilde{z})=-\int^{\tilde{z}} \partial_{\tilde{r}} \tilde{c}/\tilde{c}^2dz'$.  Then the metric \eqref{analogue} is written as
\bea
&&\hspace{-7mm}ds^2=\left.\sqrt{\frac{\rho_s}{gm}}\lambda^2\tilde{c}^2\right\{-dT^2+dZ^2+\left[\left(\frac{\lambda_\perp}{\lambda \tilde{c}}\right)^2+f^2\right]d\tilde{r}^2 \nonumber\\
&&\hspace{15mm}\left.+\left(\frac{\tilde{r}\lambda_\perp}{\lambda \tilde{c}}\right)^2d\theta^2-2f\;dZd\tilde{r} \right\}.
\eea
Note that $f$ vanishes in the limit where we can neglect the radial derivatives of $\tilde{c}$. In that limit note that the metric takes a particularly simple diagonal form.

Relabeling $\phi:=\widehat\theta_1$ the action for phase perturbations written in these coordinates is
\bea
S&=&-\frac{1}{2}\int d^4x\sqrt{\mbf{g}}\;g^{\mu\nu}\partial_\mu\phi\;\partial_\nu\phi\nonumber\\
&=& -\frac{1}{2}\int dTdZ\;\left[-\Omega(T,Z)(\partial_T\phi)^2+H(T,Z)(\partial_Z\phi)^2\right]\nonumber\\
&&\hspace{32mm}+\mathcal{O}(\partial_r\phi)+\mathcal{O}(\partial_\theta\phi) \label{fullaction}
\eea
where
\be
\Omega=\int dr d\theta \sqrt{\frac{\rho_s}{gm}}r\lambda_\perp^2\, , \label{Om}
\ee
and
\be
H=\int dr d\theta \sqrt{\frac{\rho_s}{gm}}r\lambda_\perp^2\left[1+\left(\frac{f\lambda\tilde{c}}{\lambda_\perp}\right)^2\right]\equiv \Omega+\delta\Omega. \label{aH}
\ee
The action \eqref{fullaction} has been written up to terms involving the radial and angular derivatives of the field $\phi$ which we will discard at this stage. The justification for this approximation is transparent: We  consider condensates sufficiently tightly trapped in the radial direction such that the infrared cutoff provided by the finite radial size and the ultraviolet cutoff provided by the minimal healing length in fact preclude the appearance of any radial or angular modes at all. In other words, in order for a mode of the field $\phi$ to satisfy the ultraviolet band-limitation it must necessarily not contain any non-zero radial or angular frequencies.  Or even more simply, due to the small radial size of the condensate radial and angular modes necessarily would oscillate on length scales shorter than the healing length and are hence absent.

Explicit formulas for the functions $\Omega$ and $\delta\Omega$ in terms of the lab coordinates $(t,z)$ are obtained by integrating \eqref{Om} and \eqref{aH}
\be
\Omega(t,z)=\frac{\Omega_0}{\omega_\perp^2}\frac{\lambda_\perp}{\sqrt{\lambda}}\left(1-\tilde{\omega}_{||}\frac{z^2}{\lambda^2}\right)^{3/2} \label{Omega}
\ee
and
\begin{align}
\delta\Omega(t,z)=\frac{2\Omega_0}{\omega_\perp}\frac{\lambda^{3/2}}{\lambda_\perp}&\left(\tilde{\omega}_{||}^{-1/2}\frac{z}{\lambda}-3\frac{z^2}{\lambda^2}\right. \nonumber \\
&\left.+3\tilde{\omega}_{||}^{1/2}\frac{z^3}{\lambda^3}-\tilde{\omega}_{||}\frac{z^4}{\lambda^4} \right)
\end{align}
where
\be
\Omega_0=\frac{4\pi}{3g}\left(\frac{\mu}{m}\right)^{3/2}.
\ee
Here it is understood that these functions are zero if the right hand sides become negative.  Both functions $\Omega$ and $\delta\Omega$ have support only over the extent of the condensate.  $\Omega$ monotonically decreases from its maximum at $z=0$ whereas $\delta\Omega$ vanishes at $z=0$ increases to a maxima at $\tilde{z}=\tilde{\omega}_{||}^{-1/2}/4$ before decaying back down to zero.  With the parameters we are using here the maxima of $\delta\Omega$ is approximately $\Omega_0/(\tilde{\omega}_{||}\tilde{\omega}_\perp)$ which, due to the large trapping frequencies $\omega_i$, is five orders of magnitude smaller than $\Omega_0/\omega_\perp^3$, the maximum of $\Omega$.  Hence in what follows we set $H=\Omega$, neglecting the $\delta\Omega$ contribution to $H$.

In this limit (and neglecting terms containing radial and angular derivatives, as discussed) the $1+1$ dimensional action \eqref{fullaction} becomes particularly simple
\be
S=-\frac{1}{2}\int dTdZ\;\Omega(T,Z)\left[-(\partial_T\phi)^2+(\partial_Z\phi)^2\right] ,\label{onedaction}
\ee
leading to the field equation
\be
-\partial_T\left(\Omega\partial_T\phi\right)+\partial_Z\left(\Omega\partial_Z\phi\right)=0\,. \label{nontriv}
\ee
In what follows we shall work with this field equation.

\subsection{Conformal symmetry approximation}

\begin{figure}
 \begin{center}
  \includegraphics[scale=0.4]{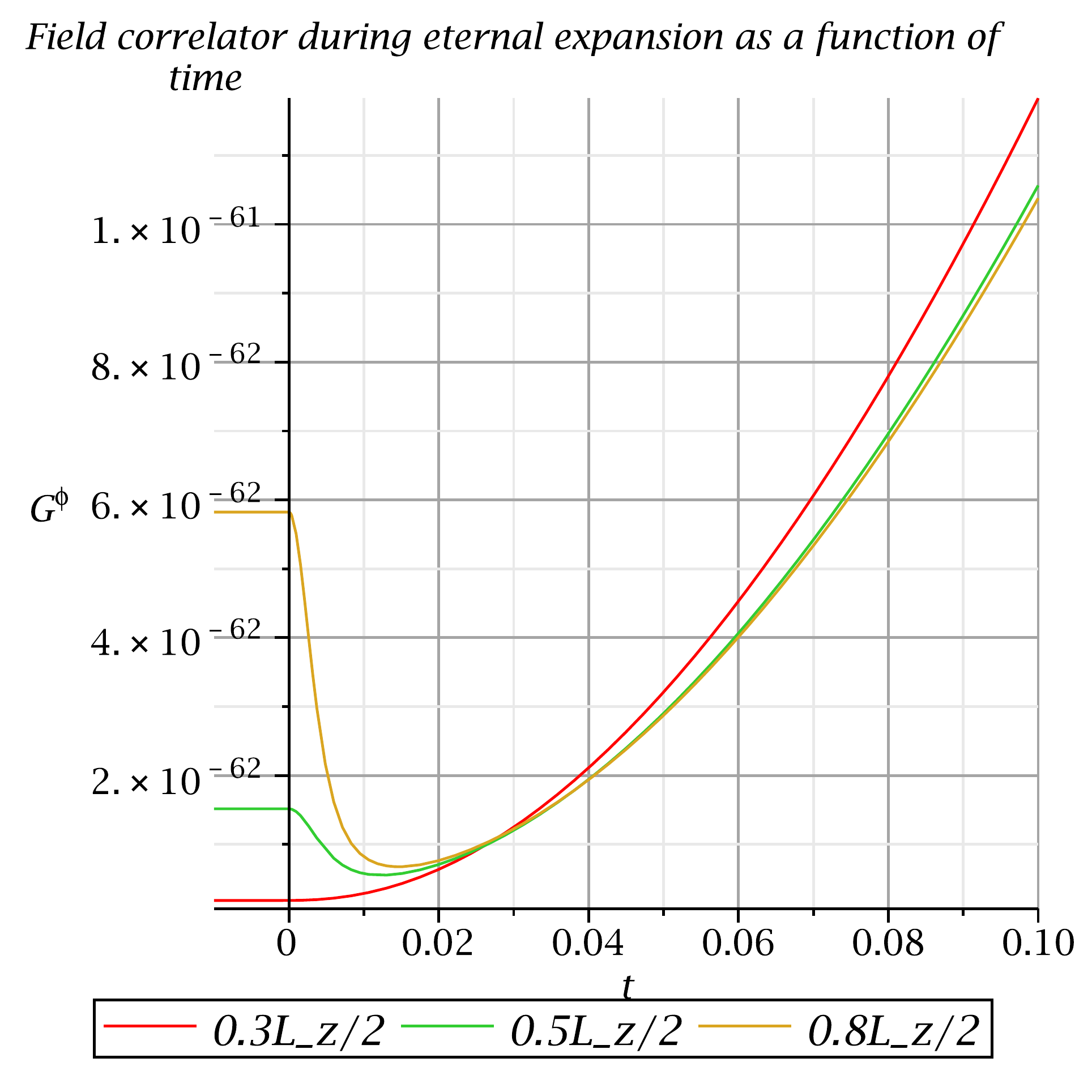}\caption{The field (phase) correlations $G^\phi$ as a function of time $t$ for three different fixed laboratory spatial separations $0.2L_z$, $0.5L_z/2$and $0.8L_z/2$. \label{trivcorrfig}}
 \end{center}
\end{figure}

As a zeroth order attempt at uncovering a signature of ``cosmological" particle creation in the Wightman function one might be tempted to discard derivatives of the function $\Omega$.  Under such an approximation the field equation reduces to the flat 1+1 dimensional wave equation for the scaled field $\tilde{\phi}=\Omega\phi$ whose solutions are simple exponentials.  The Wightman function for the phase field $\phi$ is then simply related to the standard Green function as
\begin{align}
\langle\phi(T,Z)\phi(T',Z')\rangle&=\frac{\langle\tilde{\phi}(T,Z)\tilde{\phi}(T',Z')\rangle}{\Omega(T,Z)\Omega(T',Z')}\nonumber\\
&=-\frac{\hbar}{4\pi}\;\frac{\text{ln}\,\left[\Delta x^+\Delta
x^-\right]}{\Omega(T,Z)\Omega(T',Z')} \label{trivcorr}
\end{align}
where $x^\pm$ are the characteristic null coordinates.
However this is immediately seen to be too strong an approximation to capture any particle production effects.  In fact the above correlator can be easily plotted as in Fig.~\ref{trivcorrfig}, which shows no sign of the typical transient propagating over-correlation feature normally associated with particle production (see section \ref{threeplusone}).

{This observation seems to be at odds} with several results found
in some the literature concerning the correlation structure in the
case of Hawking radiation. For example, in
\cite{Balbinot:2007de,Carusotto:2008ep} an analogous calculation is
performed to the one here, dimensionally reducing a 3+1D BEC
dynamical problem down to one in 1+1D in the case of an acoustic
black hole background geometry for phase perturbations. The
important point for the current discussion is that the authors of
\cite{Balbinot:2007de, Carusotto:2008ep} in fact do observe
numerically a signature of created Hawking quanta in the same
approximation which neglects all the derivatives of the function
$\Omega$. However, there the non-trivial ``cross-horizon"
correlations are contained entirely to the relationship between the
lab coordinates $x$ and $t$ and the null co-ordinates $x^\pm$.
Indeed also in \cite{Schutzhold:2010ig} these correlations across
the horizon are demonstrated for an explicitly 1+1D black hole
geometry\footnote{Note the difference between the analyses of
\cite{Schutzhold:2010ig} and  \cite{Balbinot:2007de,
Carusotto:2008ep}: whereas \cite{Schutzhold:2010ig} works
exclusively with the 1+1D conformally flat geometry the minimally
coupled field the 1+1D action of \cite{Balbinot:2007de,
Carusotto:2008ep} arises from a dimensional reduction of a 3+1D
model.  This dimensional reduction is encoded in the 1+1D dynamics
by the factor $\Omega$ which renders the field non-minimally coupled
to the geometry. } which, being conformally flat, possesses the
standard logarithmic functional form for the correlator.  Again the
Hawking signal is contained exclusively in the relationship between
the null and lab coordinate functions.

The reason for this discrepancy is rooted in the rather different causal structure of the black hole and cosmological analogue spacetimes and in the inherently different natures of the associated particle creation processes. In the black hole spacetime the presence of the acoustic horizon is associated to an ergoregion which allows for negative energy states (w.r.t. an asymptotic observer) and these states in turn allow for the stationary Hawking flux on the black hole static spacetime. In the case of the cosmological analogue geometry there is not such an ergoregion and particle production is permitted only as a result of the time dependence of the geometry. This time dependence is fully encoded in the conformal factor describing the difference from Minkowski space of the analogue geometry.  It is then clear that when one introduces an approximation which induces a conformal symmetry in the field equation, no particle production can take place.

\subsection{Spatially Homogeneous  $\Omega(Z)$ approximation}

As discussed above, the approximation in which we discard both temporal and spatial derivatives of $\Omega$ is too strong to capture the physics of particle creation in the 1+1 dimensional massless case. However, we shall now argue that it is sufficient to keep the $T$ derivatives while neglecting the $Z$ derivatives of $\Omega$ to observe to particle creation signal.

Firstly, since the relationship between $Z$ and the co-moving lab coordinate $\tilde{z}$ is given by integrating \eqref{exact} (in the approximation which neglects radial and angular derivatives, i.e. $f=0$) to
\be
Z(\tilde{z})=\frac{\sqrt{2}}{\omega_z}\;\text{arcsin}\left(\sqrt{\frac{m\omega_z^2}{2\mu}}\tilde{z}\right)
\ee
we see that $\partial_Z\Omega$ is naturally suppressed. One has
\be
\partial_Z\Omega=\frac{\partial z}{\partial Z}\partial_z\Omega.
\ee
where the first factor $\partial z/\partial Z$ is naturally small near the boundary (but still within the region in which the TF approximation remains valid) and the second factor $\partial_z\Omega$ is naturally small in the central region where the condensate density is almost constant.  Therefore, the factor $\partial_Z\Omega$ is naturally small everywhere on the condensate.

No such natural cancellation is available for $\partial_T\Omega$ since $\partial t/\partial T=\lambda^{3/2}\lambda_\perp $, which is not naturally a small factor.
Hence in what follows we will keep only $T$ derivatives of $\Omega$ and discard the $Z$ derivatives.

Note that this approximation is not the same as assuming $\Omega$ to be constant in lab coordinates, the difference being in the extra factor of $dz/dZ$ which suppresses $\partial_Z$ at the boundary of the condensate.   An important feature of our analysis here is the inclusion of the effects of non-constant density and finite condensate size.

\subsubsection{Phase-Phase Correlations}

Consider, then, \eqref{nontriv} and assume $\Omega$ is a function only of the variable $T$.  Then \eqref{nontriv} is written
\be
-\partial_T^2\phi-\frac{\Omega_{,T}}{\Omega}\partial_T\phi+\partial_Z^2\phi .\label{cosmoeqn}
\ee
 Let $\eta$ be defined by the differential expression $dT=\Omega\, d\eta$.  Then \eqref{cosmoeqn} separates to
\be
\partial_\eta^2\phi_{k}+\Omega^2(\eta)k^2\phi_k=0 \label{probing}
\ee
where $\phi_k=(2\pi)^{-1}\int_L dZ\,\phi \, \text{e}^{-ikZ}$ is the $k$th Fourier component of the function $\phi$ and $L$ is the $Z$-size of the condensate.

Since we work in the compact region $L$, inverse Fourier integrals will be replaced by discrete Fourier series in this section.  Hence $k$ in \eqref{probing} is constrained to satisfy $k=2n\pi/L$ after choosing the boundary conditions $\phi(-L/2)=\phi(L/2)=0$.

Such an equation of motion can be solved exactly, as we did in the previous section whenever $\Omega^2$ is contained in the three parameter family of functions
\be
\mathcal{S}:=\left\{\left.\frac{a_i^2+a_f^2}{2}+\frac{a_f^2-a_i^2}{2}\;\text{tanh}\left(\frac{\eta}{\eta_0}\right)\;\right|\;a_i,\,a_f,\,\eta_0\in \mathbb{R}\right\}
\ee
Our strategy here will be to make use of the available exact solutions and try to find a suitable element in $\mathcal{S}$ labelled by $(a_i,a_f, \eta_0)$ which most accurately reproduces the exact function $\Omega$ associated with free expansion of the BEC. We will see that such a class of functions is indeed appropriate for an approximation to the free expansion case with the key benefit of a description in terms of asymptotic particle states.

An alternative way to think about this approximation methodology is that one is studying the correlations in a BEC which really has this particular form for the function $\Omega$. As will be discussed later, there is a rather direct link between $\Omega$ and the scaling functions  $\lambda$.  In fact from \eqref{Omega} we see that $\Omega^2\propto\lambda^{-1}$ so that $\Omega^2\in \mathcal{S}$ are models for ``finite expansion" condensates (or ``finite contraction'' depending on the relative magnitudes of $a_i$ and $a_f$).  Such finite expansion (contraction) condensates are achievable in the lab with a suitable ramp-down or ramp-up trapping potential (see sec \ref{finite expansion}).  Note that the finite expansion case qualitatively differs from the free expansion case only in the future asymptotically static region.

Recall from Fig.~\ref{typical} that the typical frequency of produced particles during expansion is $\dot{\lambda}(t)/\lambda(t)$ which converges to zero at late times for a linear expansion $\lambda(t)\propto t$.  For this reason we expect all the particle production in the infinite expansion case to occur only during the ``accelerating" phase and that all of the interesting physics to have ceased by the time the scaling function becomes a linear function of time which occurs at late time. In this way we expect the finite expansion approximation to free expansion to capture the relevant physics
neglecting only the irrelevant linear phase of expansion at late times.

The strategy to understand particle production effects in the correlator is as follows.  We choose the quantum state to be $|\text{in}\rangle$ (and since we work in the Heisenberg picture the system remains in this state for all time) and express our results in the $\eta\rightarrow\infty$ limit in terms of the number eigenstates of the asymptotic Hamiltonian which are the physical particle states in that region.

Let us now move back to the $\theta$ notation for the phase perturbation field. The field operator is written in 1+1D as the sum
\be
\theta=\sum_k b_kf_k+b_k^{\dag}f_k^*
\ee
where the functions $f_k$ satisfy the equation of motion \eqref{probing} and are proportional to plane waves in the past and linear combinations of plane waves in the future as
\be
f_k(\eta\rightarrow\infty,Z)\propto \text{sin}\,kZ\left(\alpha_k\text{e}^{-i\omega_k\eta}+\beta_k\text{e}^{i\omega_k\eta}\right).
\ee
The normalization for the functions $f_k$ is not determined by \eqref{probing} but instead by the consistency between the commutation relations for $b_k$ with the commutation relations between the field operator $\theta$ and $\rho$, the density perturbation field. To fix this normalization we will use the commutator $[\theta(Z), \partial_t\theta(Z')]$.  In 3+1D in the TF approximation with zero background velocity flow in the BEC one has $\rho=-\partial_t\theta/g$ as well as the fundamental commutator $[\,\theta(x),\rho(x')\,]=-i\hbar\delta^3(x,x')$ implying $[\,\theta(x), \partial_t\theta(x')\,]=i \hbar g\delta^3(x,x')$.  Recall that we assume the perturbation field to be constant over the cross sectional area of the condensate (which was related to the existence of a UV cutoff at the healing length). Hence integrating this commutator we get for the 1+1D phase field  an extra factor of the area from $\rho^{(3+1)}A_\perp=\rho^{(1+1)}$ and hence
\be
[\,\theta(x), \partial_t\theta(x')\,]=\frac{i\hbar g}{A_\perp}\delta(x,x').
\ee
Further, expressing this commutator in $Z$ coordinates we recall that the conjugate momentum is a density of weight one whereas the field $\theta$ is a scalar so that one picks up a factor of the Jacobian under a coordinate transformation
\be
[\,\theta(Z), \partial_t\theta(Z')\,]=[\,\theta(x), \partial_t\theta(x')\,]\times \frac{dZ}{dx}
\ee
Therefore
\be
[\,\theta(Z), \partial_t\theta(Z')\,]=\frac{i\hbar g\lambda_f}{c_{s,0}A_\perp}\delta(Z,Z')
\ee
where $c_{s,0}$ is the initial sound velocity (which appears as a result of the use of $\tilde{c}_s$ instead of $c_s$ in the definition of the coordinate $Z$) and $\lambda_f$ is the final value of the scale factor $\lambda$ (which comes from the relationship between co-moving $\tilde{x}$ and lab $x$).
The dispersion relation is easily computed to be $\omega_k^2=\Omega^2k^2\rightarrow a_f^2k^2$ using the metric in $(\eta,Z)$ coordinates.
Hence we arrive at the expression
\begin{align}
\frac{ig\hbar\lambda_f}{c_{s,0} A_\perp}\delta(Z,Z')&= [\,\theta(Z), \partial_t\theta(Z')\,]\nonumber \\
& = \sum_{k,k'}\;[\,b_k, b^\dag_{k}\,]\left(f_k\partial_tf_{k}^*-f_k^*\partial_tf_{k}\right)
\end{align}
which relates the commutators and the Wronskian for the mode functions allowing us to fix the normalization factor.

For a Fock space (particle) representation we require $[\,b_k, b_{k'}^\dag\,]=\delta_{kk'}$.  Hence we arrive at the Wronskian constraint on the mode functions
\be
\sum_k\left(f_k\partial_tf_{k'}^*-f_k^*\partial_tf_{k'}\right)=\frac{ig\hbar\lambda_f}{c_{s,0} A_\perp}\delta(Z,Z').
\ee
In order to take the laboratory time derivatives above we require the relationship between $t$ and $\eta$. We have
\be
d\eta=\frac{dT}{\Omega}=\frac{1}{\Omega}\frac{dt}{\lambda^{3/2}}=\frac{\Omega^2}{\tilde{\Omega}_0^3}dt
\ee
where we have firstly transformed to the variable $T$ using $dT=\Omega d\eta$ and consequently to $t$ using \eqref{exact}, expressing the result in terms of $\Omega$ alone. $\Omega$ and $\lambda$ are related by  \eqref{Omega} as $\Omega=\tilde{\Omega}_0/\sqrt{\lambda}$ where $\tilde{\Omega}_0$ is the spatial part of $\Omega$, numerically $\tilde{\Omega}_0\approx 1.5\times 10^{37}$. Then
\be
dt=\frac{\tilde{\Omega}_0^3}{\Omega^2}\,d\eta=\frac{\tilde{\Omega}_0^3\; d\eta}{\frac{\textstyle a_i^2+a_f^2}{\textstyle 2}+\frac{\textstyle a_f^2-a_i^2}{\textstyle  2}\;\text{tanh}\left(\frac{\textstyle \eta}{\textstyle \eta_0}\right)}\label{times}
\ee

Now, we wish to approximate the free expansion scale factor shown in Fig.~\ref{scale} for approximately the first $t=0.15$s. This choice of the approximation region is motivated by two constraints: Firstly, after about $t=0.15$s, a large proportion of the condensate no longer is described accurately by a TF approximation as can be seen in Fig.~\ref{quantumratio} ; Secondly the natural shape of the exact free expansion scaling function is less accurately fit by the tanh functions we use here over longer time periods since the scaling function enters a linear regime at late times.

Over this $t=0.15$s time interval the scaling function $\lambda$ increases from $\lambda_i=1$ to approximately $\lambda_f=25$.  Since $\Omega=\tilde{\Omega}_0/\sqrt{\lambda}$  we see that the variable $\Omega$ must \emph{decrease} from $\tilde{\Omega}_0$ to $\tilde{\Omega}_0/5$. Therefore we must have $a_f=a_i/5$.
Hence by \eqref{times} we have the asymptotic relations between the time variables
\be
t(\eta)=\left\{\begin{array}l \lambda_i\tilde{\Omega}_0\,\eta,\quad\text{for}\; t\rightarrow-\infty \vspace{1mm}\\
\lambda_f\tilde{\Omega}_0\,\eta,\quad\text{for}\; t\rightarrow \infty
\end{array}\right.
\ee
where the function $t(\eta)$, obtained by integrating exactly the differential expression \eqref{times}
, varies smoothly between these two constant asymptotes in the intermediate region.  Taking the average gradient in the intermediate region for $t(\eta)$ we see that a $t$ interval of $0.15$s corresponds to an $\eta$ interval of approximately $12\times\tilde{\Omega}_0\approx8.2\times 10^{-40}$ (in units of $[\Omega^{-1}]$s or [Energy]$\cdot$[Time]$^2$).
Therefore, we require the function $\Omega(\eta)$ to vary over this time scale between the initial and final values $a_i$ and $a_f$. That is we should choose
\be
\eta_0\approx 8.2\times 10^{-40}\;\text{[Energy]}\cdot\text{[Time]}^2
\ee
With these choices one finds that $\omega_k\eta=kt/\lambda_f^{3/2}$.

For the un-normalized $f_k$ we have
\be
\sum_k \left(f_k\partial_tf_{k}^*-f_k^*\partial_tf_{k}\right)=2\sum_k\frac{ik}{\lambda_f^{3/2}}\;\text{sin}kZ\;\text{sin}kZ'.
\ee
Scaling the functions $f_k$ as
\be
\tilde{f}_k=\sqrt{\frac{\lambda_f^{3/2} \hbar g}{\text{Vol}\,A_\perp c_s^i k}}f_k,
\ee
where Vol is the axial Z-length of the condensate (which has the dimension of a time), the Wronskian provides the correct numerical and functional form for the commutator.

Hence the correctly normalized mode functions are given by
\be
f_k=\frac{\sqrt{\lambda_f^{3/2}\hbar g}}{\sqrt{c_s^iA_\perp \text{Vol}}}\frac{\text{sin}(kZ)}{k}\;\text{e}^{-ikt/\lambda_f^{3/2}}
\ee
Here we have transformed back to lab $t$ coordinates in which we take the time derivatives.

In 1+1D at late times the equal time Wightman function is written in terms of $\alpha_k$ and $\beta_k$ in a very similar way to the 3+1D example above in terms of the discrete (and inhomogeneous) Fourier series
\begin{align}
G^\theta(\eta, Z,Z')=&N\displaystyle\sum\limits_{n=1}^{\infty}\frac{1}{n}\;\text{sin}\left(k_n Z\right)\text{sin}\left(k_n Z'\right)\nonumber\\
&\times\left(1+2|\beta_{k_n}|^2+2\,\mathfrak{Re}\,\alpha_{k_n}\beta_{k_n}^*\text{e}^{-2ik_na_f\eta}\right) \label{diswight}
\end{align}
where the normalization factor N is given by
\be
N=\frac{\lambda_f^{3/2}\hbar g}{A_\perp c_{s,0} \pi}.
\ee
Taking symmetrically spaced points from the center of the condensate, the transient time dependent contribution is given by
\begin{align}
\mathcal{I}=N\displaystyle\sum\limits_{n=1}^{\infty}\frac{1}{n}\;\text{sin}&\left[n\pi \left(1+\frac{2Z}{L}\right)\right]\text{sin}\left[n\pi\left(1-\frac{2Z}{L}\right)\right]\nonumber\\
&\times2\,\mathfrak{Re}\,\alpha_{k_n}\beta_{k_n}^*\text{e}^{-2 ik_na_f\eta} \label{1+1Ent}
\end{align}
is easily computed as a truncated sum.

\begin{figure}
 \begin{center}
  \includegraphics[scale=0.4]{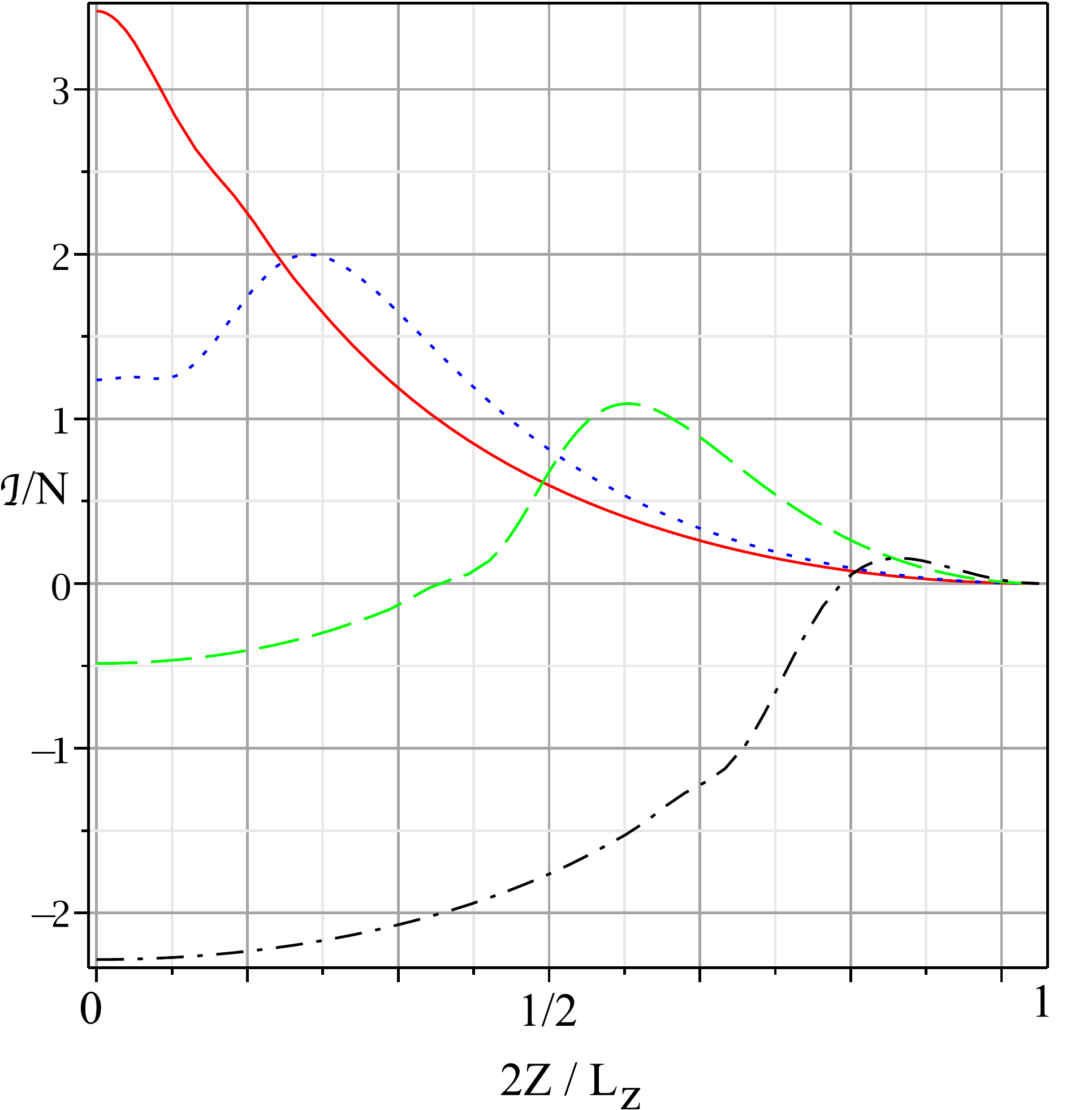} \caption{The bump of correlations propagating outwards from the center of the BEC cloud. The curves are for $t=0$s (red, solid), $t=0.3$s (blue, dotted), $t=0.8$s (green dashed) and $t=1.2$s (black, dash dots) Here we plot the truncated sum up to the first $90$ terms. The horizontal axis of this plot is the fraction of the half length $L_z/2$ of the condensate after expansion. The vertical axis is the integrated contribution Ent in units of the normalization factor $N$.    \label{discrete1+1bump}}
 \end{center}
\end{figure}

With the choices made above for the parameters $a_i$, $a_f$ and $\eta_0$ we find the entanglement structure shown in Fig. \ref{discrete1+1bump}.
The time scales of the propagating bump are understood in terms of the crossing time for the massless modes: After the expansion the sound speed has decreased to one fifth its initial value $c_s/5$ due to the dilution of the BEC gas ($\lambda(t\rightarrow\infty)=25$). The size of the condensate has increased by a factor of $25$ so the crossing time increases by a factor of $125$. Initially with a axial length of $L_z=5\times 10^{-5}$m and $c_s=2\times10^{-3}$ms$^{-1}$ the crossing time is $t_{\text{crossing}}=5\times10^{-2}$s which increases to the order of seconds after expansion. (Note again as in the 3+1D case, that the entanglement propagation speed is twice the sound speed since it represents pairs of phonons traveling in opposite directions).  This is a slight overestimation since the sound speed actually decreases from a maximum at the center of the condensate to zero at the edge. However we have neglected this slowing in the analysis.    This order of magnitude estimate is confirmed in the plot of Fig. \ref{discrete1+1bump}.

\begin{figure}[htb]
\begin{center}$
\begin{array}{c}
\includegraphics[scale=0.35]{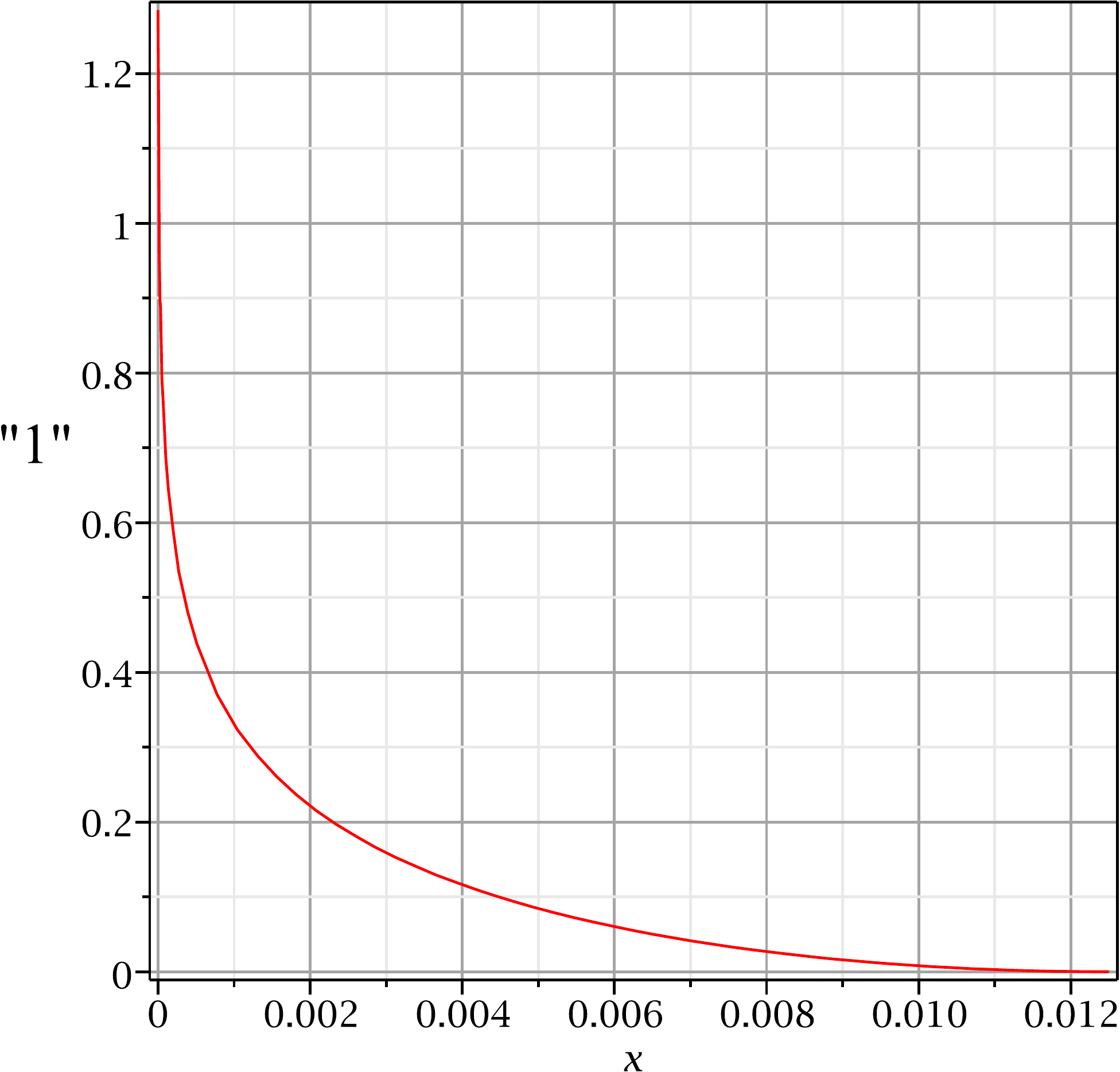} \\
\includegraphics[scale=0.35]{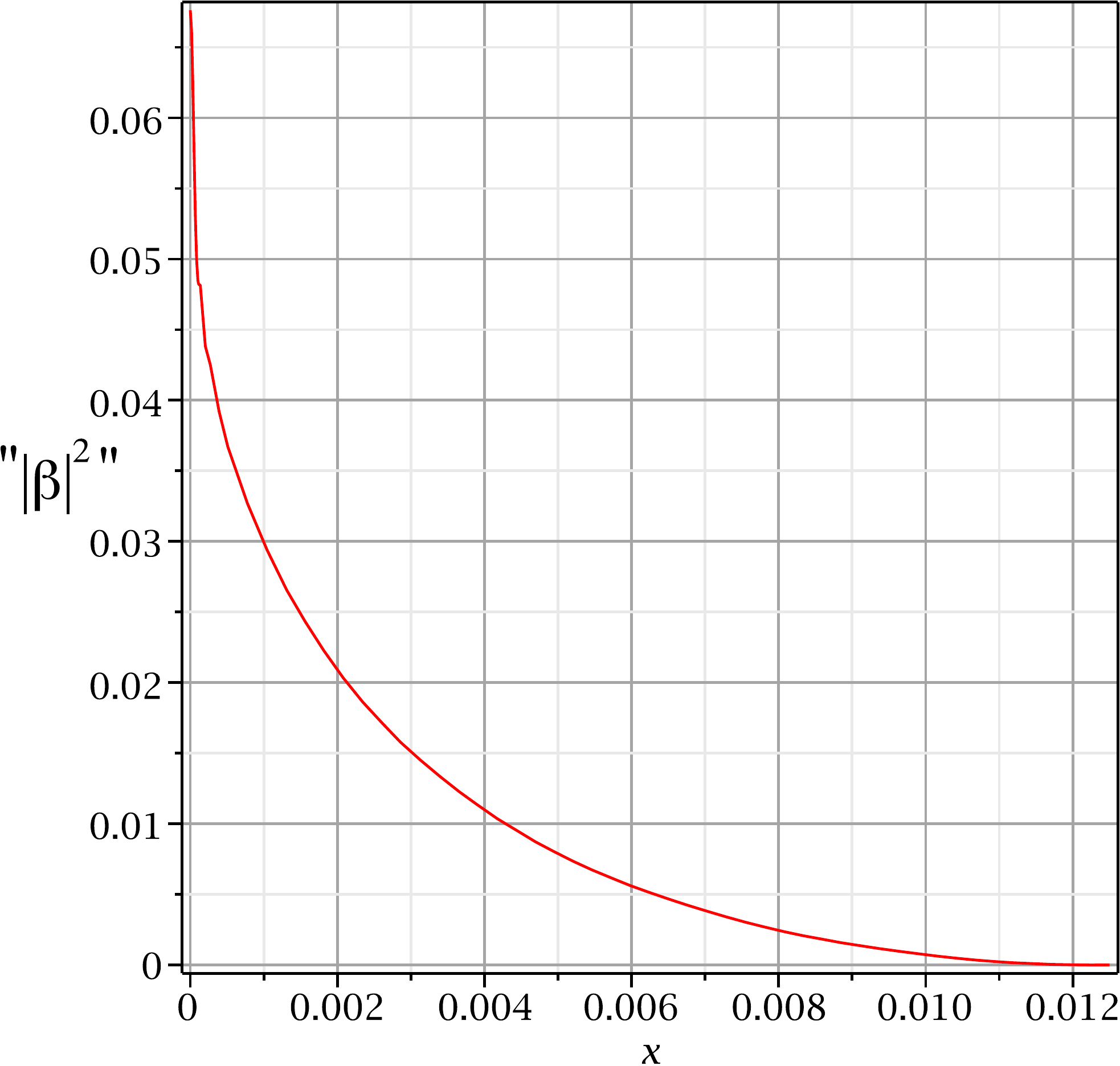}
\end{array}$
\caption{The contributions to the phase Wightman function from the ``1'' and ``$|\beta_k|^2$'' terms in \eqref{diswight} truncated to the first 200 term of the series. Such a truncation serves as an ultraviolet regulator on the correlation singularity at the origin of the ``1'' background contribution.  No such UV divergence is present in the ``$|\beta_k|^2$''  contribution. \label{relative}}
\end{center}
\end{figure}

In Fig.~\ref{relative}
we show also the integrated contributions from the ``$1$'' and ``$|\beta_k|^2$'' terms in the integrand, in the same units as Fig.~\ref{discrete1+1bump}.  We note that the magnitude of the bump structure is in general much larger than the background (``1'') and enhancement (``$|\beta_k|^2$'') contributions.  This is expected since the produced particles propagate from small distances where there is a high intrinsic correlation to long distances where the natural background fluctuations are no longer correlated.

%
%
%

%

\subsubsection{Density-Density Correlations}

\begin{figure}[htb]
\begin{center}$
\begin{array}{c}
\includegraphics[scale=0.32]{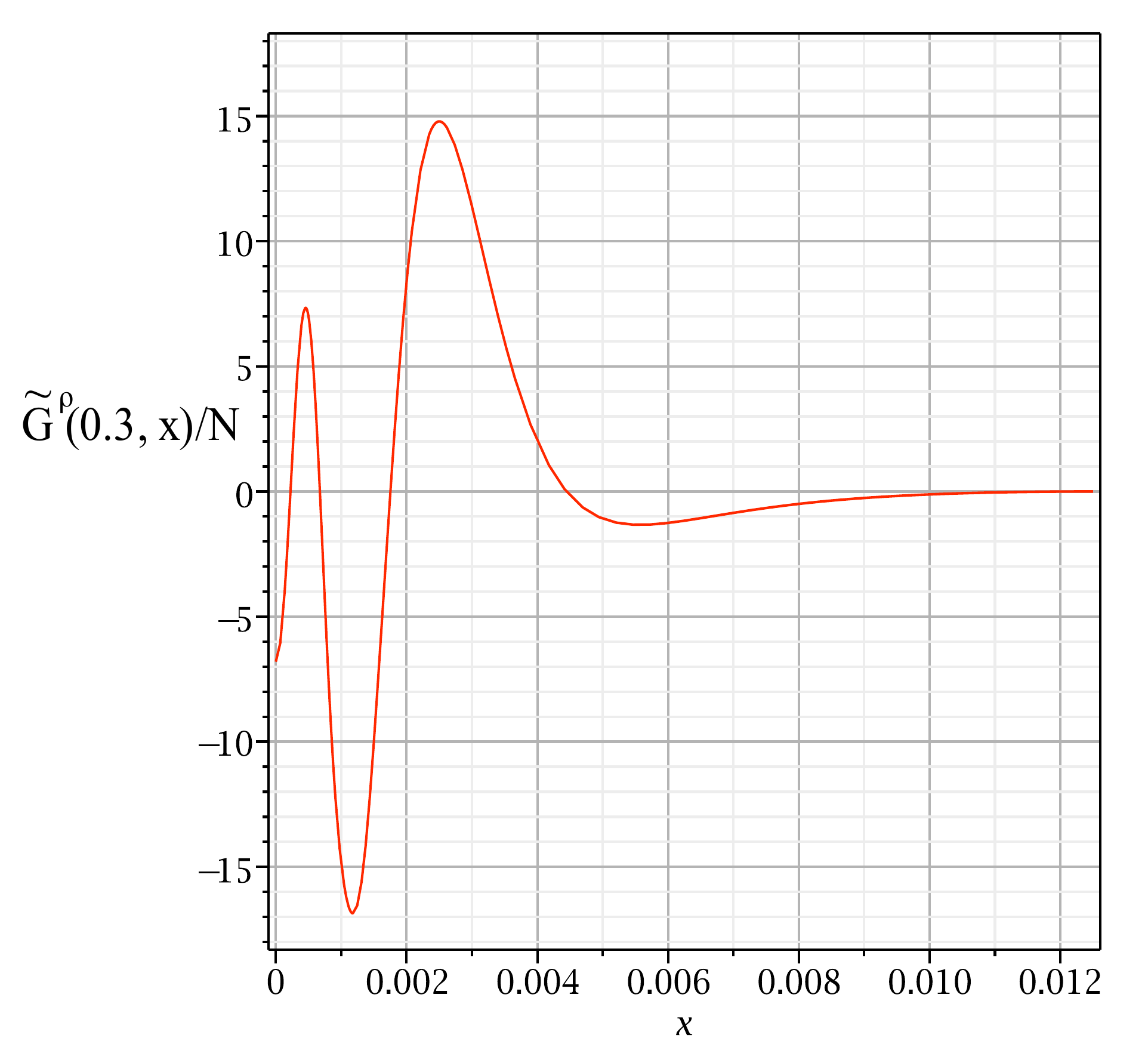} \\
\includegraphics[scale=0.32]{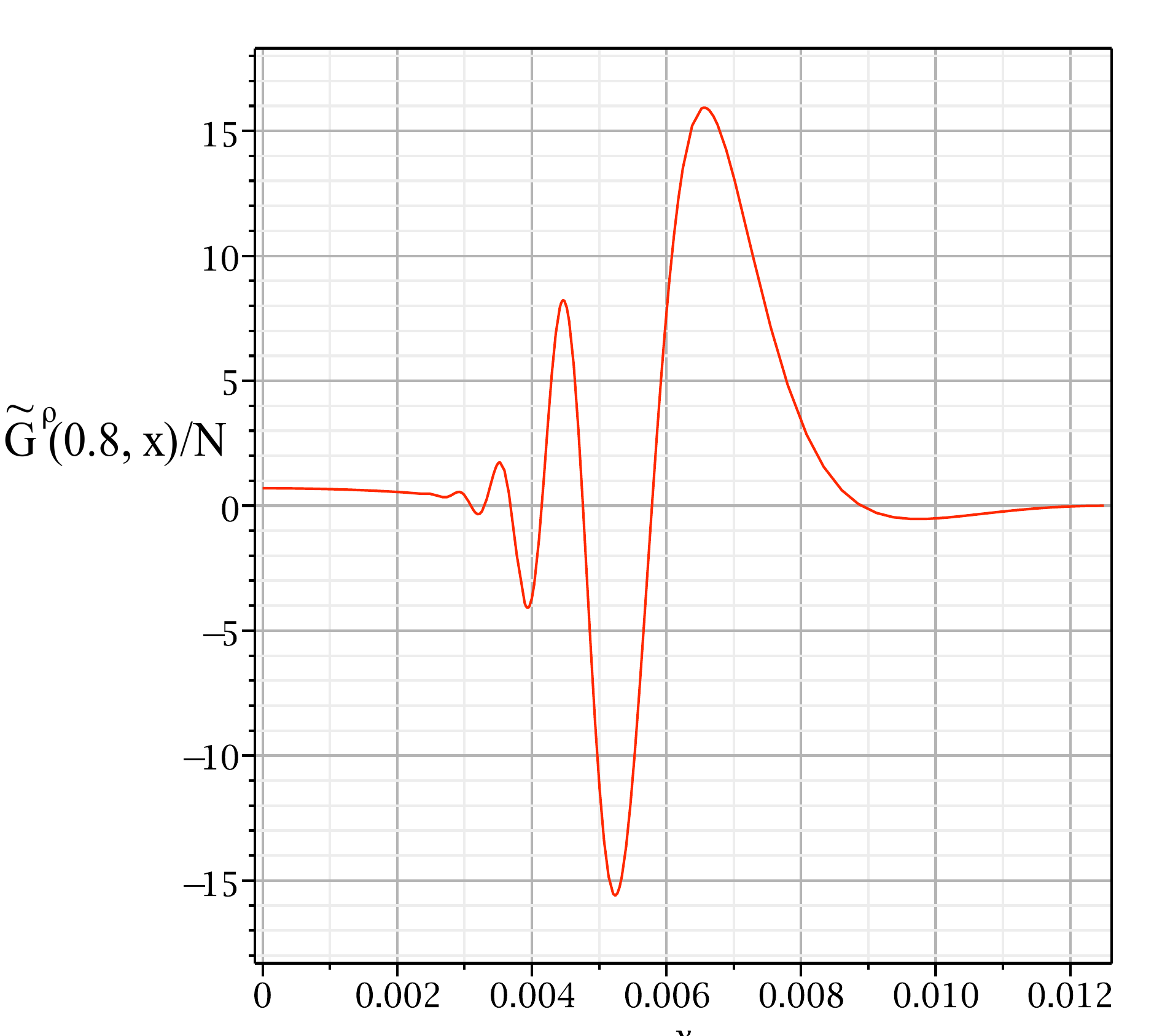} \\
\includegraphics[scale=0.32]{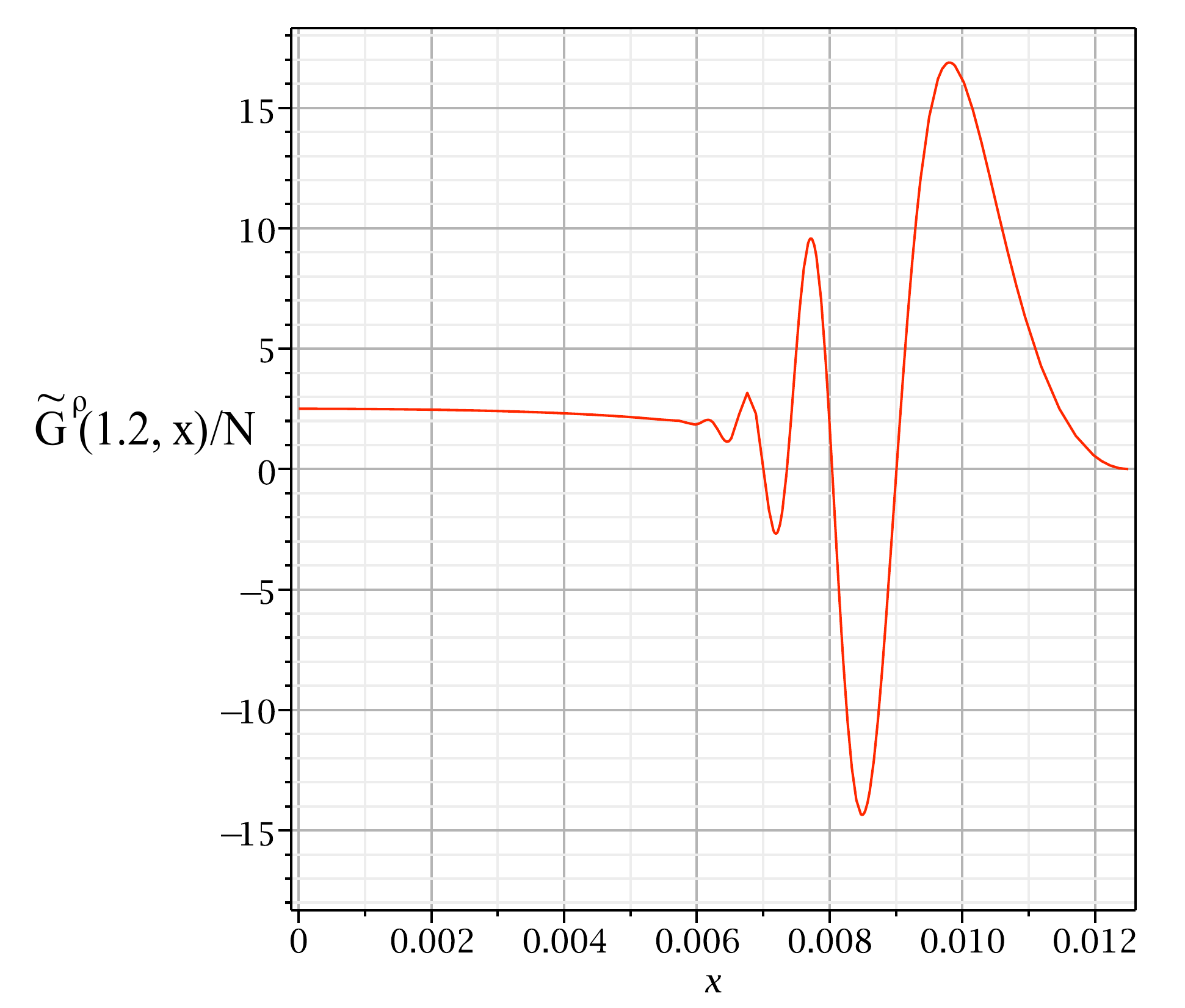}
\end{array}$
\caption{The density correlation function after expansion has taken place, for the finite expansion case, in units of $N\times A_\perp^2/g^2$.. The curves from top to bottom correspond to $t=0.3$s, $t=0.8$s and $t=1.2$s after the onset of expansion (which we take, as described above, to last for $0.15$s).  \label{mom}}
\end{center}
\end{figure}

Experimentally, although there are proposals to directly measure the phase correlations,
the density correlations are those of major observational interest.
Again, in the TF approximation the density field is written in terms of the phase as
\be
\rho^{(3+1)}=-\frac{1}{g}\partial_t\theta-\frac{1}{g}v_z\partial_z\theta.
\ee
The 1+1D density is related to the 3+1D density by $\rho^{(3+1)}=\rho^{(1+1)}A_\perp$ where $A_\perp$ is the cross sectional area of the condensate (which we assume constant).   Then the density correlator is derived according to
\be
G^{\rho}(z,z')=\text{lim}_{t\rightarrow t'}\mathcal{D}G^\theta(t,t',z,z')
\ee
where we define the differential operator $\mathcal{D}$ as
\be
\mathcal{D}=\frac{A_\perp^2}{g^2}\left(\partial_t+v_z\partial_z\right)\left(\partial_{t'}+v_{z'}\partial_{z'}\right).
\ee
In the 1+1D finite expansion case, $\mathcal{D}$ reduces to simply the time derivative since $v_z=0$ after the expansion has finished. In Fig. \ref{mom} we plot the normalized  density correlator $\tilde{G}^\rho(t,z,z'):=G^\rho(t,z,z')/(\rho^{(1+1)})^2$ for symmetrically placed points $z, z'$ about the middle of the condensate in units of $N\times A_\perp^2/g^2$.
We note again the propagating structure but of a slightly more complicated shape to the phase correlation bump.

\subsection{Finite expansion of an elongated BEC \label{finite expansion}}

\begin{figure}
 \begin{center}
  \includegraphics[scale=0.4]{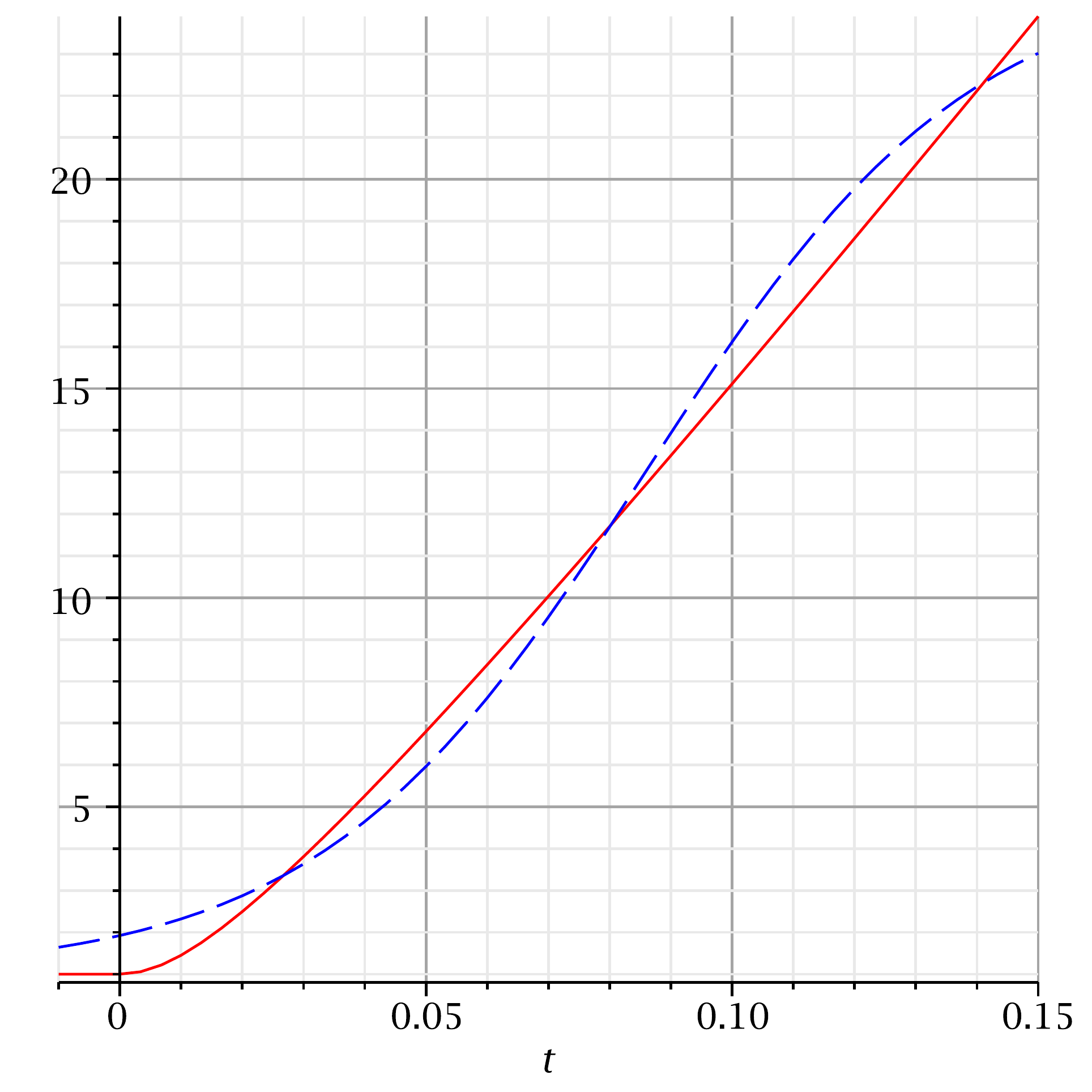}\caption{The exact scaling function $\lambda$ (red, solid) and its approximation by the tanh function $\tilde{\Omega}_0^2/\Omega$ (blue, dashed) over the interval $t=0.15$s as described in the text.  \label{finiteapproxtoeternal}}
 \end{center}
\end{figure}

The availability of an asymptotically static regime in the future is a necessary ingredient for the application of the Bogoliubov formalism. As discussed above there are two ways to use the tools contained in that formalism to understand a system which does not possess such a future static regime such as the case of indefinite expansion of a released BEC.  Either one considers the finite expansion as a theoretical approximation to the ``true" unconstrained expansion case, as shown in Fig.~\ref{finiteapproxtoeternal}
or alternatively one can imagine actually performing a finite expansion experiment in the lab with a condensate which is at first released and consequently ``caught'' after a finite expansion. Above we have essentially been discussing the first interpretation.  Here we briefly discuss the second.

In order for $\lambda(t)$ to follow the correct step up profile with a future non-dynamical region, one might expect a simple two level potential would suffice. However, the equations of motion \eqref{scalingf} for the scaling function admit oscillating solutions so that in order to arrive at a static condensate in the asymptotic future a rather finely tuned potential is necessary.

Again assuming the constancy of the radial scaling functions during the axial expansion we have for the trapping frequency
\be
\omega_{||}^2(t)=\frac{\omega_{||}^2(0)}{\lambda^3(t)}-\frac{\ddot{\lambda}(t)}{\lambda(t)}.
\ee
The necessary trapping frequency is obtained by inserting the desired expansion profile into this expression.  One finds that in fact, the squared trapping frequency becomes negative for a short time near the end of the transition from expansion to staticity indicating a temporarily repulsive force is necessary to stop the BEC from contracting further.

In this way it would be possible to actually create in the lab a BEC system with particular time dependent trapping frequencies such that the scaling function $\lambda$ behaves in the way appropriate (as discussed above) to induce the kind of finite expansion or contraction analogue geometry for perturbations.  It is worth pointing out again at this point that we expect the general structure of a propagating bump of correlations resulting from a finite scaling to not depend in any crucial way on the fine details of the scaling. We expect the result to be insensitive for example to the fact that a precise reproduction of the tanh shape profile is practically very difficult and most probably would only be approximated in any real experiment.

\section{Conclusion}

In this paper we have considered the correlation structure of expanding BECs both in the idealized case of isotropic and homogeneous expansion as well as in the experimentally relevant case of a finite size, cigar-shaped, BEC anisotropically expanding.

In the first case we have developed a  general formalism for probing the effects of particle production on correlations as embodied in the equal-time Wightman function. Specifically, for any spatially translation invariant system we have shown that it makes sense to define the correlator as in \eqref{Wight} which in the case of particle production (using Bogoliubov coefficients) takes the form \eqref{int}.  This result is completely generic to particle production in any spatially translation invariant system, applying to standard cosmologies as much as the ``analogue spacetimes" of central interest in the current article. Specifically, the general cosmological result was shown to be directly implementable in a 3+1D BEC system with a time dependent
scattering length as well as a 3+1D isotropically expanding BEC, opening two alternative window on possible observation of this correlation signal.

In the second case (anisotropic BEC) we have shown that the condensate dynamics reduces to an effectively $1+1$-dimensional system and that the Wightman function can be generically cast in the form \eqref{diswight} for these systems. Using this result we have studied the correlation structure and shown how it carries the signature of analogue cosmological particle production possibly observable in future experiments. In particular however, we have studied the density correlations in such BEC systems, which are of current experimental interest.

The magnitude of the normalized correlations is of pertinent experimental concern. The overall scale of the propagating bump is given in terms of the normalization factor for phase perturbation along with the additional factors which relate the 1+1D density to the phase field
$\frac{\langle\widehat\rho\widehat\rho\rangle}{\rho^2}\simeq\frac{N}{g^2}\frac{1}{(\rho^{(1+1)})^2}\approx
10^{-7}$.
Although this is a very small number, it should be noted that this
magnitude is in fact much larger than the magnitude of the
background correlation structure (present in flat spacetime also).
The relative magnitude of the background to entanglement
correlations is independent of the normalization of the mode function solutions but instead depends on the structure of the Bogoliubov
coefficients  which derive from the wave equation alone.

The main purpose of this paper was to investigate the characteristic
signature related to the cosmological particle emission in an expanding
(or more generally time-dependent) BEC, and analyze it without
specific attention to its actual measurability.  {However, one can of course imagine
altering} the parameters of the background BEC in such a way to
improve the signal. For example, one parameter which can be altered
by an order of magnitude is the number of atoms $N$ which affects
the value of $\rho_0$ entering into the denominator of the
normalized correlations. Another possibility of improving the signal
would be to alter the scattering length. We have discussed above the
use of the analogue FRW geometry when one induces a time dependence
in the scattering length. Here we have something different in mind.
Since the background density depends on the scattering length as
$a^{3/5}$, by decreasing and holding the scattering length constant
before expansion, one might be able to improve the signal.   Another
possibility is to choose a longer time scale on which to
approximate the eternal expansion with a finite expansion; By
stopping the expansion later, the correlations will be propagating
on a background density which is lower which would improve the ratio
in the normalized correlations. All these ways of altering $\rho_0$
will also affect the values of $\eta_0$, $a_i$ and $a_f$ possibly
giving a more pronounced signal.



A separate analysis should regard the role of a non-vanishing
initial temperature of the BEC (or more generally the role of a non-vacuum initial state). In fact, as already pointed out in \cite{Carusotto:2008ep, Carusotto:2009re, PhysRevA.81.031610},
the net effect of such an initial temperature (i.e.~a non ``empty"
initial state, but thermally populated) is to significantly increase
the quantity of created excitations, therefore enhancing the signal
without affecting the mechanism responsible for the particle
production. We intend to make further investigations in these directions in future work.


\begin{acknowledgments}
The authors wish to thank Matt Visser for useful comments and suggestions.
\end{acknowledgments}
\bibliography{pig}

\end{document}